\documentclass[letterpaper]{article} 
\usepackage{aaai18}  %Required
\usepackage{times}  %Required
\usepackage{helvet}  %Required
\usepackage{courier}  %Required
\usepackage{url}  %Required
\usepackage{graphicx}  %Required
\usepackage{amssymb}
\usepackage{amsmath}
\usepackage{amsthm}
\usepackage{thmtools,thm-restate}
\usepackage{pifont}
\usepackage{color}
\usepackage{xspace}
\usepackage{misc}
\newcommand{\ELHI}{\ensuremath{\mathcal{ELHI}}\xspace}

\newif\ifextended
\extendedtrue

\newtheorem{theorem}{Theorem}
\newtheorem{lemma}[theorem]{Lemma}
\newtheorem{definition}[theorem]{Definition}
\newtheorem{example}[theorem]{Example}
\newtheorem{corollary}[theorem]{Corollary}

\begin{document}

\title{Query Expressibility and Verification in Ontology-Based Data Access}
\author{Carsten Lutz, Johannes Marti, Leif Sabellek\\
Department of Computer Science, University of Bremen, Germany
}
\maketitle
\begin{abstract}
  In ontology-based data access, multiple data sources are integrated
  using an ontology and mappings. In practice, this is often achieved
  by a bootstrapping process, that is, the ontology and mappings are
  first designed to support only the most important queries over the
  sources and then gradually extended to enable additional queries.
  In this paper, we study two reasoning problems that support such
  an approach. The expressibility problem asks whether a given source
  query $q_s$ is expressible as a target query (that is, over the
  ontology's vocabulary) and the verification problem asks,
  additionally given a candidate target query $q_t$, whether $q_t$
  expresses $q_s$.  We consider (U)CQs as source and target queries
  and GAV mappings, showing that both problems are $\Pi^p_2$-complete
  in DL-Lite, \coNExpTime-complete between \EL and \ELHI when
  source queries are rooted, and 2\ExpTime-complete for unrestricted
  source queries.
\end{abstract}

\section{Introduction}

Ontology-based data access (OBDA) \cite{poggi-2008} is an
instantiation of the classical data integration scenario, that is, a set
of data sources is translated into a unifying global schema by means
of mappings.
%  which enables the formulation of queries over the global
% schema.
% as well as their automatic translation back into queries over
% the sources
The distinguishing feature of OBDA is that the global schema is
formulated in terms of an ontology which provides a rich domain model
and can be used to derive additional query answers via logical
reasoning. % For some concrete industrial use
% cases of OBDA, we refer to
% \cite{DBLP:conf/semweb/KharlamovHJLLPR15,DBLP:conf/semweb/KharlamovBGJKLM16}.
%
When data sources are numerous such as in large enterprises, data
integration is often a considerable investment. OBDA is no exception
since the construction of both the mappings and the ontology is
non-trivial and labour intensive. 

In practice, OBDA is thus often approached in an incremental manner
\cite{DBLP:conf/sigmod/TrisoliniLN99,DBLP:conf/semweb/KharlamovHJLLPR15,DBLP:journals/internet/SequedaM17}. One
starts with a small set of important source queries (typically hand
crafted by experts from the enterprise's IT
department) % that require multiple
% sources to be answered
and builds mappings for the involved sources and an initial ontology
that support these queries, manually or with the help of extraction
tools \cite{DBLP:conf/semweb/Jimenez-RuizKZH15,sickshit}. The outcome
of this first step is then evaluated and, when considered successful,
ontology and mappings are extended to support additional queries.
This process may proceed for several rounds and in fact forever since
new data sources and queries tend to appear as the enterprise develops
and existing data sources or the ontology need to be updated
\cite{DBLP:conf/ijcai/LemboRSST17}.

The aim of this paper is to study two reasoning tasks that support
such an incremental approach to OBDA. The \emph{expressibility
  problem} asks whether a given source query $q_s$ is expressible as a
target query $q_t$ over the global schema defined by the ontology. Possible
reasons for non-expressibility include that the mappings do not 
transport all data required for answering $q_s$ to the global schema
and that the ontology `blurs' the distinction between different
relations from the sources, examples are given in the
paper.  If $q_s$ is not expressible, one might thus decide to add
more mappings or to rework the ontology. The verification problem
asks, additionally given a candidate target query $q_t$, whether $q_t$
expresses $q_s$. This is useful for example when a complex $q_t$ has
been manually constructed and when the ontology, mappings, or source
schemas have been updated, with an unclear impact on $q_t$. The same
problems have been considered in the context of open data publishing,
there called finding and recognition of s-to-t rewritings
\cite{DBLP:conf/dlog/Cima17}.

We consider UCQs (and sometimes CQs) both for source and target
queries, global as view (GAV) mappings, and ontologies that are
formulated in DL-Lite or in a description logic (DL) between \EL and
\ELHI, which are all very common choices in OBDA. It follows from
results in
\cite{DBLP:journals/tods/NashSV10,DBLP:journals/tcs/Afrati11} that,
even without ontologies, additional source UCQs become expressible
when full first-order logic (FO) is admitted for the target query
rather than only UCQs. In OBDA, however, going beyond UCQs quickly
results in undecidability of query answering
\cite{DBLP:books/daglib/0041477} and thus we stick with UCQs.

The expressibility problem in OBDA is closely related to the problem
of query expressibility over views, which has been intensively studied
in database theory, see for example
\cite{Levy95,DBLP:conf/pods/DuschkaG97,DBLP:conf/pods/CalvaneseGLV02,DBLP:journals/tods/NashSV10,DBLP:journals/tcs/Afrati11}
and references therein. The
problem has occasionally also been considered in a DL context
\cite{DBLP:conf/aaai/CalvaneseGL00,DBLP:conf/ihis/HaaseM05,DBLP:conf/pods/BeeriLR97,DBLP:journals/jcss/CalvaneseGLR12}. These
papers, however, study setups different from the one we consider, both
regarding the r\^ole of the ontology and the description logics used.

In many classical cases of query expressibility over views, informally
stated, $q_s$ is expressible over a set of mappings \Mbf (representing
views) if and only if the UCQ $\Mbf^-(\Mbf(q_s))$ is contained in
$q_s$ where $\Mbf(q_s)$ is the UCQ obtained from $q_s$ by applying the
mappings and $\Mbf^-(\Mbf(q_s))$ is obtained from $\Mbf(q_s)$ by
applying the mappings backwards
\cite{DBLP:journals/tods/NashSV10,DBLP:journals/tcs/Afrati11}. Our
starting point for proving decidability and upper complexity bounds
for expressibility in OBDA is the observation that we need to check
whether $\Mbf^-(q_r)$ is contained in $q_s$ where $q_r$ is a
(potentially infinitary) UCQ-rewriting of the UCQ $\Mbf(q_s)$ under
the ontology; note that, here, we mean \emph{rewriting} of an
ontology-mediated query into a source query in the classical sense of
ontology-mediated querying, see for example
\cite{BHLW-IJCAI16}. Verification can be characterized in a very
similar way. These characterizations also show that expressibility can
be reduced to verification in polynomial time and that if $q_s$ is
expressible, then it is expressed by the polynomial size UCQ
$\Mbf(q_s)$.

Our main results are that within the setup described above,
expressibility and verification are $\Pi^p_2$-complete in
DL-Lite$^\Rmc_{\mn{horn}}$ and in many other dialects of DL-Lite, {\sc
  co\-NExpTime}-complete in DLs between \EL and \ELHI when the source
UCQ is rooted (that is, every variable is reachable from an answer
variable in the query graph of every CQ), and 2\ExpTime-complete in
the unrestricted case. There are some surprises here. First, the
$\Pi^p_2$ lower bound already applies when the ontology is empty and
the source query is a CQ which means that, in the database theory
setting, it is $\Pi^p_2$-hard to decide the fundamental problem
whether a source CQ is expressible as a (U)CQ over a set of UCQ
views. For this problem, an \NPclass upper bound was claimed without
proof in \cite{Levy95}. Our results show that the problem is actually
$\Pi^p_2$-complete. A second surprise is that 2\ExpTime- respectively
\coNExpTime-hardness applies already in the case that the ontology is
formulated in \EL (and when queries are UCQs).  We are not aware of
any other reasoning problem for \EL that has such a high complexity
whereas there are several such problems known for \ELI, that is, \EL
extended with inverse roles \cite{BHLW-IJCAI16}. There is a clear
explanation, though: the mappings make it possible to introduce just
enough inverse roles in the backwards translation $\Mbf^-$ mentioned
above so that hardness proofs can be made work.

\ifextended
Detailed proofs are deferred to the appendix.
\else
Detailed proofs are deferred to the appendix that is included in the
extended version of the paper available at
\url{http://www.informatik.uni-bremen.de/tdki/research/papers.html}.
\fi

\section{Preliminaries}

We use a mix of standard DL notation
\cite{DBLP:books/daglib/0041477} and standard notation from database
theory.

% We introduce databases and queries, query containment, homomorphisms,
% the DLs relevant for this paper, GAV mappings, as well as OBDA
% specifications and the main reasoning problems studied.

\emph{Databases and Queries}.  A \emph{schema} \Sbf is a set of
relation names with associated arities. An \emph{\Sbf-database} $D$ is
a set of \emph{facts} $R(a_1,\dots,a_n)$ where $R \in \Sbf$ is a
relation name of arity $n$ and $a_1,\dots,a_n$ are constants.  We use
$\mn{adom}(D)$ to denote the set of constants that occur in $D$.
% An \emph{\Sbf-ABox} is a database over DL schema
% \Sbf. An interpretation \Imc is a \emph{model} of an ABox \Amc if
% $A(a) \in \Amc$ implies $a \in A^\Imc$ and $r(a,b) \in \Amc$ implies
% $(a,b) \in r^\Imc$. In contrast to standard DL terminology, we speak
% of constants and facts also in the context of ABoxes, instead of
% individuals and assertions.
%
% Note that $\mn{adom}(\Amc)$ denotes the set of all
% constants that occur in the ABox \Amc, often called individuals in a
% DL context.

\emph{A conjunctive query (CQ) $q(\xbf)$ over schema $\Sbf$} takes the
form $\exists \ybf \, \varphi(\xbf,\ybf)$, where $\xbf$ are the
\emph{answer variables}, \ybf are the \emph{quantified
  variables}, and $\varphi$ is a conjunction of \emph{relational
  atoms} $R(z_{1},\ldots,z_{n})$ and \emph{equality atoms} $z_1 = z_2$
where $R\in \Sbf$ is of arity $n$ and $z_{1},\ldots,z_{n}$ are variables
from $\xbf \cup \ybf$. Contrary to the usual setup and to avoid
dealing with special cases in some technical constructions, we do
\emph{not} require that all variables in \xbf actually occur in
$\varphi(\xbf,\ybf)$.  We sometimes confuse $q$ with the set of atoms
in $\varphi$, writing for example $R(x,y,z) \in q$. We use
$\mn{var}(q)$ to denote $\xbf \cup \ybf$.  The \emph{arity} of a CQ is
the number of variables in \xbf and $q$ is \emph{Boolean} if it has
arity
zero. % We remark that $q(\xbf)$ can be regarded as a database $D_q$,
% the \emph{canonical database of $q(\xbf)$}, whose facts are exactly
% the relational atoms in $\varphi(\xbf,\ybf)$ with variables viewed as
% constants. 
A
\emph{homomorphism from $q$ to a database $D$} is a function
$h: \mn{var}(q) \rightarrow \mn{adom}(D)$ such that $R(h(\xbf)) \in D$
for every relational atom $R(\xbf) \in q$ and $h(x)=h(y)$ for
every relational atom $x=y \in q$. Note that $h$ needs to be
defined also for answer variables that do not occur in
$\vp(\xbf,\ybf)$. A tuple $\abf \in \mn{adom}(D)$ is an \emph{answer}
to $q$ on $D$ if there is a homomorphism $h$ from $q$ to
$D$ with $h(\xbf)=\abf$. A \emph{union of conjunctive queries (UCQ)}
is a disjunction of CQs that all have the same answer
variables. Answers to UCQs are defined in the expected way. We use
$\mn{ans}_{q}(D)$ to denote the set of all answers to UCQ
$q$ on database $D$.

Let $q_1(\xbf_1), q_2(\xbf_2)$ be UCQs of the same arity and over the
same schema \Sbf. We say that $q_1$ is \emph{contained} in~$q_2$,
denoted $q_1 \subseteq_\Sbf q_2$, if for every \Sbf-database $D$,
$\mn{ans}_{q_1}(D) \subseteq \mn{ans}_{q_2}(D)$. It is well-known
that, when $q_1,q_2$ are CQs, then $q_1 \subseteq_\Sbf q_2$ iff there
is a homomorphism from $q_2$ to $q_1$, that is, a function
$h: \mn{var}(q_2) \rightarrow \mn{var}(q_1)$ such that
$R(h(\xbf)) \in q_1$ for every relational atom $R(\xbf) \in q_2$,
$(h(x),h(y))$ is in the equivalence relation generated by the
equality atoms in $q_1$ for every $x=y \in q_2$,
and $h(\xbf_2)=\xbf_1$. We indicate the existence of
such a homomorphism with $q_2 \rightarrow q_1$. When $q_i$ is a UCQ
with disjuncts $q_{i,1},\dots,q_{i,k_i}$, $i \in \{1,2\}$, then
$q_1 \subseteq_\Sbf q_2$ iff for every $q_{1,i}$, there is a $q_{2,j}$
with $q_{2,j} \rightarrow q_{1,i}$.

We shall frequently view CQs as databases, for which we merely need to
read variables as constants of the same name and drop equality
atoms. Conversely, we shall also view a tuple $(D,\abf)$ with $D$ a
database and $\abf=a_1 \cdots a_n \in \mn{adom}(D)$ as an $n$-ary CQ;
note that repeated elements are admitted in \abf. We do this by
reserving $n$ fresh answer variables $x_1,\dots,x_n$, viewing $D$ as a
CQ by reading all constants (including those in \abf) as quantified
variables, and adding the equality atom $x_i=a_i$ for $1 \leq i
\leq n$.

\medskip

\emph{Ontology-Based Data Access}.  
Let \NC and \NR be
countably infinite set of \emph{concept names} and \emph{role
  names}. An \emph{\ELI-concept} is formed according to the syntax
rule
$$
  C,D ::= \top \mid A \mid C \sqcap D \mid \exists r . C  \mid \exists r^- . C
$$
where $A$ ranges over \emph{concept name} and $r$ over \emph{role names}.  An
expression $r^-$ is called an \emph{inverse role} and a \emph{role} is
either a role name or an inverse role. As usual, we let $(r^-)^-$
denote $r$. An \emph{\ELHI-ontology} is a finite set of \emph{concept
  inclusions (CIs)} $C \sqsubseteq D$, $C,D$ \ELI-concepts, and
\emph{role inclusions} $r \sqsubseteq s$ and $r \sqsubseteq s^-$. The
semantics is defined in terms of interpretation
$\Imc=(\Delta^\Imc,\cdot^\Imc)$ as usual. An \emph{\EL-concept} is an
\ELI-concept that does not use the constructor $\exists r^- .C$. An
\emph{\EL-ontology} is an \ELHI-ontology that uses only \EL-concepts
and contains no role inclusions. A \emph{basic concept} is a concept
name or of one of the forms $\top$, $\bot$, $\exists r . \top$, and
$\exists r^- . \top$. A DL-Lite$^\Rmc_{\mn{horn}}$-ontology is a
finite set of statements of the form
$$
   B_1 \sqcap \cdots \sqcap B_n \sqsubseteq B 
   \quad 
   r \sqsubseteq s 
   \quad 
   r \sqsubseteq s^-
   \quad
   r_1 \sqcap \cdots \sqcap r_n \sqsubseteq \bot 
$$
where $B_1,\dots,B_n,B$ range over basic concepts and
$r,s,r_1,\dots,r_n$ range over role names.

% An ABox
% \Amc is \emph{satisfiable} w.r.t.\ an ontology \Omc iff \Amc and \Omc
% have a common model.

A \emph{DL schema} is a schema that uses only unary and binary
relation names, which we identify with \NC and \NR respectively.  An
\emph{\Sbf-ABox} is a database over DL schema \Sbf. An interpretation
\Imc is a \emph{model} of an ABox \Amc if $A(a) \in \Amc$ implies $a
\in A^\Imc$ and $r(a,b) \in \Amc$ implies $(a,b) \in r^\Imc$. In
contrast to standard DL terminology, we speak of constants and facts
also in the context of ABoxes, instead of individuals and assertions.

An \emph{ontology-mediated query (OMQ)} is a tuple $Q=(\Omc,\Sbf,q)$
with \Omc an ontology, \Sbf a DL schema, and $q$ a query such as a CQ.
Let \Amc be an \Sbf-ABox. A tuple $\abf \in \mn{adom}(\Amc)$ is a
\emph{certain answer to $Q$ on \Amc} if $\abf \in \mn{ans}_q(\Imc)$
for every model \Imc of \Amc (viewed as a potentially infinite
\Sbf-database).  We use $\mn{cert}_Q(\Amc)$ to denote the set of all
certain answers to $Q$ on~\Amc and sometimes write $\Amc,\Omc \models
q(\abf)$ when $\abf \in \mn{cert}_Q(\Amc)$. For OMQs
$Q_1=(\Omc_1,\Sbf,q_1)$ and $Q_2=(\Omc_2,\Sbf,q_2)$ of the same
arity, we say that $Q_1$ is \emph{contained} in $Q_2$, denoted $Q_1
\subseteq Q_2$, if for every $\Sbf$-ABox \Amc,
$\mn{cert}_{Q_1}(\Amc) \subseteq \mn{cert}_{Q_2}(\Amc)$. Containment
between an OMQ and a UCQ are defined in the expected way, and so is
the converse containment.

A \emph{global as view (GAV) mapping} over a schema \Sbf takes the
form $\vp(\xbf,\ybf) \rightarrow \psi(\xbf)$ where $\vp(\xbf,\ybf)$ is a
conjunction of relational atoms over \Sbf and $\psi(\xbf)$ is of the
form $A(x)$, $r(x,y)$, or $r(x,x)$ with $A$ a concept name and $r$ a
role name.
% \footnote{The results in this paper also hold with equality
%   atoms admitted in $\vp(\xbf)$, but we omit them here for
%   simplicity.} 
We call $\vp(\xbf,\ybf)$ the \emph{body} of the mapping and
$\psi(\xbf)$ its \emph{head}. Every variable that occurs in the head must
also occur in the body. Let \Mbf be a set of GAV mappings over a
schema~\Sbf. For every \Sbf-database~$D$, the mappings in $\Mbf$
produce an ABox $M(D)$, defined as follows:
$$
%  \Mbf(D)=
\{ R(\abf) \mid D \models \varphi(\abf,\bbf) \text{ and } 
 \varphi(\xbf,\ybf) \rightarrow R(\xbf) \in \Mbf \}.
$$
This ABox can be physically materialized or left virtual; we do not
make any assumptions regarding this issue. 

An \emph{OBDA specification} is a triple $\Smc=(\Omc,\Mbf,\Sbf)$ where
$\Sbf$ is the \emph{source schema}, $\Mbf$ a finite set of mappings
over $\Sbf$, and \Omc an ontology.\footnote{For readability, we
  consider a single data source, only. Multiple source databases can
  be represented as a single one by assuming that their schemas are
  disjoint and taking the
  union.} %that does not use relation names from
%$\Sbf$.  
We use $\mn{sch}(\Mbf)$ to denote the schema that consists of all
relation names that occur in the heads of mappings in
$\Mbf$. Informally, \Smc is addressing source data in schema \Sbf,
translated into an ABox in schema $\mn{sch}(\Mbf)$ in terms of the
mappings from \Mbf and then evaluated under the ontology \Omc. Note that
\Omc can use the relation names in $\mn{sch}(\Mbf)$ as well as
additional concept and role names, and so can queries that are 
posed against the ABox.

%
% If $q$ is a CQ over $\Sbf$, we will also interpret $q$ as a \Sbf-database $I_q$ and write $\Mbf(q)$ for $\Mbf(I_q)$. We can interpret $\Mbf(q)$ as a query with the same answer variables as $q$.
%
We use $[\Lmc,\Mmc]$ to denote the set of all OBDA specifications
$(\Omc,\Mbf,\Sbf)$ where \Omc is formulated in the ontology language
\Lmc and all mappings in \Mbf are formulated in the mapping language
\Mmc and call $[\Lmc,\Mmc]$ an \emph{OBDA language}.  An example of an
OBDA language is $[\ELHI,\text{GAV}]$. In this paper, we shall
concentrate on GAV mappings. While other types of mappings such as LAV
and GLAV are also interesting \cite{poggi-2008,DBLP:conf/dlog/Cima17},
they are outside the scope of this paper.
\begin{samepage}
\begin{definition}
\label{def:main}
Let $\Qmc_s$ and $\Qmc_t$ be query languages and $[\Lmc,\Mmc]$ an OBDA
language. 
\begin{enumerate}

\item The \emph{$\Qmc_s$-to-$\Qmc_t$ verification problem in
    $[\Lmc,\Mmc]$} is to decide, given an OBDA specification
  $\Smc=(\Omc,\Mbf,\Sbf) \in [\Lmc,\Mmc]$, a source query $q_s \in
  \Qmc_s$, and a target query $q_t \in \Qmc_t$ of the same arity,
  whether $q_t$ is a \emph{realization of $q_s$ in} \Smc, that is,
  whether $\mn{ans}_{q_s}(D)=\mn{cert}_{Q}(\Mbf(D))$ for all
  $\Sbf$-databases~$D$% such that the ABox $\Mbf(D)$ is satisfiable
%  w.r.t.\ \Omc
, where $Q=(\Omc,\mn{sch}(\Mbf),q_t)$.

\item The \emph{$\Qmc_s$-to-$\Qmc_t$ expressibility problem in
  $[\Lmc,\Mmc]$} is to decide, given an OBDA specification
  $\Smc=(\Omc,\Mbf,\Sbf) \in [\Lmc,\Mmc]$ and a source query
  $q_s \in \Qmc_s$, whether there is a realization $q_t$
  of $q_s$ in $\Qmc_t$. We then say that $q_s$ is
  \emph{$\Qmc_t$-expressible in} \Smc.

\end{enumerate}
\end{definition}
\end{samepage}
Note that an alternative definition is obtained by quantifying only over
those \Sbf-databases $D$ such that $D \cup \Omc$ is satisfiable.  This
does not make a difference for most of the setups studied in this paper
since the involved DLs cannot express inconsistency.
%We close this section with some examples.
%
\begin{example}
  Assume that \Smc  contains a binary relation $\mn{Man}$ with
  $\mn{Man}(m, d)$ meaning that department $d$ is managed by manager
  $m$ and a ternary relation $\mn{Emp}(e,d,o)$ meaning that employee
  $e$ works for department $d$ in office $o$. Let \Mbf contain the 
  GAV mappings 
  $$
  \begin{array}{rcl}
    \mn{Man}(x,z) \wedge \mn{Emp}(y,z,u)
    &\rightarrow&\mn{manages}(x,y) \\[1mm]
%    \mn{Man}(x,y) & \rightarrow & \mn{Manager}(x) \\[1mm]
    \mn{Emp}(x,y,z) & \rightarrow & \mn{Employee}(x)
  \end{array}
  $$
  Then the source query $q_s(x)=\exists y \, \mn{Man}(x,y)$ is not
  expressible because the mappings do not provide
  sufficient data from the source. It trivially becomes expressible as
  $q_t(x)=\mn{Manager}(x)$ when we add the mapping
  $$
  \mn{Man}(x,y) \rightarrow \mn{Manager}(x).
  $$
  Next, we further add the following \EL-ontology \Omc:
  $$
  \begin{array}{rcl}
  \mn{Manager} &\sqsubseteq& \mn{Employee} \\[1mm]
  \mn{Manager} & \sqsubseteq& \exists \mn{manages} . \mn{Secretary}
  \end{array}
  $$
  Then the source query $q_s(x)=\exists y \exists z \,
  \mn{Emp}(x,y,z)$, which formerly was expressible as
  $q_t(x)=\mn{Employee}(x)$, is no longer expressible due to the first
  CI in \Omc. Informally, all the required data is there, but it is
  mixed with other data and we have no way to separate. The source
  query $q_s(x,y)=\exists z \exists u \, \mn{Man}(x,z) \wedge
  \mn{Emp}(y,z,u)$, however, is expressible as $\mn{manages}(x,y)$
  despite the second CI in \Omc, intuitively because the additional
  data mixed into \mn{manages} by that CI always involves an anonymous
  constant introduced through the existential quantifier and is thus
  never returned as a certain answer.
\end{example}
% We remark that is is usual in OBDA to consider only ABoxes that are
% satisfiable with the ontology, see for example
% \cite{some,stuff}. Given an OBDA specification $\Smc=(\Omc,\Mbf,\Sbf)$
% and an $\Sbf$-database $D$, it is easy to check whether $\Mbf(D)$ is
% satisfiable w.r.t.\ \Omc since $\Mbf(D)$ can be computed in polynomial
% time and satisfiability of ABoxes w.r.t.\ ontologies is a standard
% reasoning problem \cite{textbook}.
%
The \emph{size} of any syntactic object $X$ such as a
  UCQ or an ontology, denoted $|X|$, is the number of symbols needed
  to write it, with names of concepts, roles, variables, etc.\ counting
  as one. 

\section{Characterizations}

We characterize when a UCQ $q_t$ over $\mn{sch}(\Mbf)$ is a
realization of a UCQ $q_s$ over the source schema \Sbf and then lift
this characterization to the expressibility of $q_s$. This serves as a
basis for deciding the expressibility and verification problems later
on. The characterization applies the mappings from \Mbf forwards and
backwards, as also done in query rewriting under views
\cite{DBLP:journals/tods/NashSV10,DBLP:journals/tcs/Afrati11}, and
suitably mixes in UCQ-rewritings of certain emerging OMQs.

\smallskip

Let $\Smc=(\Omc,\Mbf,\Sbf)$ be an OBDA specification and
$Q=(\Omc,\mn{sch}(\Mbf),q)$ an OMQ. A \emph{rewriting} of $Q$ is a query
$q_r$ over $\mn{sch}(\Mbf)$ of the same arity as $q$ such that for all
$\mn{sch}(\Mbf)$-ABoxes \Amc, $\mn{ans}_{q_r}(\Amc)=
\mn{cert}_Q(\Amc)$. We speak of a \emph{UCQ rewriting} if $q_r$ is a
UCQ, of an \emph{infinitary UCQ rewriting} if $q_r$ is a potentially
infinite UCQ, and so on. Note that there always exists a
\emph{canonical infinitary UCQ rewriting} that is obtained by taking
all $\mn{sch}(\Mbf)$-ABoxes \Amc and answers $\abf %=a_1 \cdots a_n 
\in \mn{cert}_Q(\Amc)$ and including $(\Amc,\abf)$ viewed as a CQ as a
disjunct. This even holds when \Omc is formulated in FO without
equality. In fact, this follows from the definition of rewritings and
the fact that OMQs with \Omc formulated in FO without equality are
preserved under homomorphisms \cite{DBLP:journals/tods/BienvenuCLW14}.

\smallskip

Let $\Smc=(\Omc,\Mbf,\Sbf)$ be an OBDA specification and $\Amc$ an
ABox that uses only concept and role names from $\mn{sch}(\Mbf)$. We
say that a mapping $\vp(\xbf,\ybf) \rightarrow \psi(\xbf)$ from $\Mbf$
is \emph{suitable} for a fact $\alpha \in \Amc$ if $\psi(\xbf)$ and
$\alpha$ are unifiable. We write $\Mbf^-(\Amc)$ to denote the set of
\Sbf-databases obtained from $\Amc$ as follows: for every fact $\alpha
\in \Amc$, choose a suitable mapping $\vp(\xbf,\ybf) \rightarrow
\psi(\xbf)$ from $\Mbf$ and include $R(\sigma(\zbf))$ in
$\Mbf^-(\Amc)$ whenever $R(\zbf)$ is an atom in $\vp(\xbf,\ybf)$ and
where $\sigma$ is the most general unifier of $\psi(\xbf)$ and
$\alpha$, extended to replace every variable from $\ybf$ with a fresh
constant. For example, for a fact $r(a,a) \in \Amc$ we can choose a
mapping $R(x,y,z) \rightarrow r(x,y)$ and include $R(a,a,b)$ in
$\Mbf^-(\Amc)$, where $b$ is fresh. Both \Mbf and $\Mbf^-$ lift to
sets of databases and ABoxes as expected, that is, if $S$ is a set of
\Sbf-databases, then $\Mbf(S)=\{ \Mbf(D) \mid D \in S\}$ and if $S$ is
a set of ABoxes over $\mn{sch}(\Mbf)$, then $\Mbf^-(S)=\bigcup_{\Amc
  \in S} \Mbf^-(\Amc)$.

% Let $\Smc=(\Omc,\Mbf,\Sbf)$ be an OBDA specification and $\Amc$ an
% ABox that uses only concept and role names from $\mn{sch}(\Mbf)$. We
% write $\Mbf^-(\Amc)$ to denote the set of \Sbf-databases obtained from
% $\Amc$ as follows: for every assertion $\alpha \in \Amc$, choose a
% mapping $\vp(\xbf) \rightarrow \psi(\ybf)$ from $\Mbf$ and include
% $R(\sigma(\zbf))$ in $\Mbf^-(\Amc)$ whenever $R(\zbf)$ is an atom in
% $\vp(\xbf)$ and where $\sigma$ is the most general unifier of $\alpha$
% and $\psi(\ybf)$, extended to replace every variable that is not in
% $\ybf$ with a fresh constant. For example, for an assertion $r(a,a)
% \in \Amc$ we can choose a mapping $R(x,y,z) \rightarrow r(x,y)$
% and include $R(a,a,b)$ in $\Mbf^-(\Amc)$, where $b$ is fresh.
% Both \Mbf and $\Mbf^-$ lift to sets of databases and ABoxes as
% expected, that is, if $S$ is a set of \Sbf-databases, then $\Mbf(S)=\{
% \Mbf(D) \mid D \in S\}$ and if $S$ is a set of ABoxes over
% $\mn{sch}(\Mbf)$, then $\Mbf^-(S)=\bigcup_{\Amc \in S} \Mbf^-(\Amc)$.

\smallskip

In what follows, we shall often apply \Mbf to a CQ $q(\xbf)$ viewed as
a database, and view the result (which formally is a database) again
as a CQ. In this case, the answer variables are again \xbf and the
equality atoms from $q(\xbf)$ are readded to $\Mbf(q)$. The same
applies to UCQs and sets of databases, and to $\Mbf^-(q)$ (where
we also preserve answer variables and readd equality atoms). Note
that $\Mbf^-(q)$ gives a UCQ even when $q$ was a CQ.
\begin{example}
\label{ex:main}
  Consider the CQ
  $q(x,y,z)=\exists u \, r(x,y) \wedge s(x,z) \wedge s(z,u) \wedge x=y$ and let
  \Mbf consist of the single mapping $r(x,y) \rightarrow r(x,y)$, that
  is, the role name $r$ is simply copied and the role name $s$ is
  dropped. Then $\Mbf(q)$ viewed as a CQ is 
  $p(x,y,z)=r(x,y) \wedge x=y$. Note that the answer variable $z$ does
  not occur in an atom.
\end{example}
%
% {\color{red}Do we ever use Points~(i) and~(ii)? Yes,
%   Point~(ii) is used in Theorem~\ref{thm:maincharact}.}
%
The following fundamental lemma describes the (non)-effect of applying \Mbf
and $\Mbf^-$ on query containment. It is explicit or implicit in many
papers concerned with query rewriting under views or with query 
determinacy, see for
example~\cite{DBLP:journals/tods/NashSV10,DBLP:journals/tcs/Afrati11}.
\begin{restatable}{lemma}{lempatterns}
\label{lem:patterns}
Let \Mbf be a set of GAV mappings, $q$, $q_1$ and $q_2$ UCQs over
$\Sbf$ and $r$, $r_1$ and $r_2$ UCQs over $\mn{sch}(\Mbf)$.  Then:
\begin{enumerate}
\item If $q_1 \subseteq_{\Sbf} q_2$, then $\Mbf(q_1) \subseteq_{\mn{sch}(\Mbf)} \Mbf(q_2)$.
\item If $r_1 \subseteq_{\mn{sch}(\Mbf)} r_2$, then $\Mbf^-(r_1)
  \subseteq_{\Sbf} \Mbf^-(r_2)$.
\item $q \subseteq_{\Sbf} \Mbf^-(r)$ iff $\Mbf(q) \subseteq_{\mn{sch}(\Mbf)} r$.
\end{enumerate}
\end{restatable}
The next theorem characterizes realizations in terms of
UCQ rewritings and~$\Mbf^-$. It can thus serve as a 
basis for deciding the verification problem.
\begin{theorem} \label{thm:realization} Let $\Smc=(\Omc,\Mbf,\Sbf)$ be
  an OBDA specification from $(\mbox{\text{FO}(=)},\text{GAV})$, $q_s$
  a UCQ over \Sbf, $q_t$ a UCQ over $\mn{sch}(\Mbf)$, and $q_r$ an
  infinitary UCQ rewriting of the OMQ $Q=(\Omc,\mn{sch}(\Mbf),q_t)$.
  Then $q_t$ is a realization of $q_s$ iff $q_s \equiv_{\Sbf}
  \Mbf^-(q_r)$.
\end{theorem}
\noindent
\begin{proof}
  ``if''. Assume that $q_s \equiv_{\Sbf}\Mbf^-(q_r)$. We have to show
  that $q_t$ is a realization of $q_s$. Since $q_r$ is a rewriting of
  $Q$, it suffices to prove that
  $\mn{ans}_{q_s}(D) = \mn{ans}_{q_r}(\Mbf(D))$ for all
  $\Sbf$-databases $D$.

% The original argument it is not quite correct. It does not suffice to
% show that $q_r$ is equivalent to $M(q_s)$.

%$q_r$ is equivalent to $q_s$ in the sense that $q_r \equiv M(q_s)$. From the assumption $M^-(q_r) \equiv_{\Sbf} q_s$ and Lemma \ref{lem:patterns} (iii) it follows that $M(q_s) \subseteq q_r$. For the other inclusion, let $\abf \in \mn{cert}_{\Omc, M(D)}(q_t)$. We need to show that $\abf \in \mn{ans}_{q_s}(D)$. Since $q_r$ is a rewriting of $(\Omc, q_t)$, we have $\abf \in \mn{ans}_{q_r}(M(D))$. By Lemma \ref{lem:patterns} (ii) it follows that $\abf \in \mn{ans}_{M^-(q_r)}(M^-(M(D))$ and by the assumption $M^-(q_r) \equiv_{\Sbf} q_s$ we get $\abf \in \mn{ans}_{q_s}(M^-(M(D))$. It follows from Lemma \ref{lem:patterns} (iii) that $M^-(M(D))$ always maps homomorphically into $D$, hence $\abf \in \mn{ans}_{q_s}(D)$, which completes the proof.

  For ``$\subseteq$'', assume that $\abf \in \mn{ans}_{q_s}(D)$. Let
  $p$ be $(D,\abf)$ viewed as a CQ. From $\abf \in \mn{ans}_{q_s}(D)$,
  we obtain $p \subseteq q_s$, and $q_s \subseteq_\Sbf \Mbf^-(q_r)$
  yields $p \subseteq_\Sbf \Mbf^-(q_r)$.  With Point~3 of
  Lemma~\ref{lem:patterns}, it follows that
  $\Mbf(p) \subseteq_\Sbf q_r$, which by construction of $p$ implies
  $\abf \in \mn{ans}_{q_r}(\Mbf(D))$.

  For ``$\supseteq$'', assume that $\abf \in \mn{ans}_{q_r}(\Mbf(D))$.
  Let $p$ be $(\Mbf(D),\abf)$ viewed as a~UCQ. Then,
  $p \subseteq_{\mn{sch}(\Mbf)} q_r$ and Point~3 of Lemma~\ref{lem:patterns}
  yields $p' \subseteq_\Sbf \Mbf^-(q_r)$ where $p'$ is $(D,\abf)$ viewed as a
  CQ. Together with $\Mbf^-(q_r) \subseteq_\Sbf q_s$, we obtain
  $p' \subseteq_\Sbf q_s$, which implies that $\abf \in \mn{ans}_{q_s}(D)$.

  \smallskip For the ``only if'' direction, assume that $q_t$ is a
  realization of $q_s$.  We have to show that $q_s \equiv_\Sbf
  \Mbf^-(q_r)$. Thus, let $D$ be an \Sbf-database and \abf a tuple
  from $\mn{adom}(D)$ whose length matches the arity of
  $q_s$. Further, let $p$ be $(D,\abf)$ viewed as a database. Since
  $q_t$ is a realization of $q_s$ and $q_r$ a rewriting of the OMQ
  $Q$, $\abf \in \mn{ans}_{q_s}(D)$ iff $\abf \in
  \mn{ans}_{q_r}(\Mbf(D))$. The latter is the case iff $\Mbf(p)
  \subseteq_{\mn{sch}(\Mbf)} q_r$ which by Point~3 of
  Lemma~\ref{lem:patterns} holds iff $p \subseteq_\Sbf \Mbf^-(q_r)$.
  This in turn is the case iff $\abf \in \mn{ans}_{\Mbf^-(q_r)}(D)$.
%
% ``only if''.  Assume that $q_t$ is a realization of
%   $q_s$.  We first show $q_s \subseteq_\Sbf \Mbf^-(q_r)$. Thus let
%   $\abf \in \mn{ans}_{q_s}(D)$. Since $q_t$ is a realization of $q_s$
%   and $q_r$ a rewriting of the OMQ $Q$, this means $\abf \in
%   \mn{ans}_{q_r}(\Mbf(D))$. Let $p$ be $(D,\abf)$ viewed as a
%   database. Then $\Mbf(p) \subseteq_{\mn{sch}(\Mbf)} q_r$ and Point~3
%   of Lemma~\ref{lem:patterns} entails $p \squbseteq_\Sbf \Mbf^-(q_r)$,
%   which gives $\abf \in \mn{ans}_{q_r}(\Mbf(D))$ 
%   as desired.
%
%   Now for $\Mbf^-(q_r) \subseteq_\sbf q_s$. Let $\abf \in
%   \mn{ans}_{\Mbf^-(q_r)}(D)$, $D$ an \Sbf-database. Further, let $p$
%   be $(D,\abf)$ viewed as a CQ. Then we have $p \subseteq_\Sbf
%   \Mbf^-(q_r)$.  Point~3 of Lemma~\ref{lem:patterns} yields $\Mbf(p)
%   \subseteq_{\mn{sch}(\Mbf)} q_r$. By construction of $p$ and
%   $\Mbf(p)$, we obtain $\abf \in \mn{ans}_{q_r}(\Mbf(D))$. Since $q_t$
%   is a realization of $q_s$ and $q_r$ a rewriting of the OMQ $Q$,
%   this implies $\abf \in \mn{ans}_{q_s}(D)$.
\end{proof}
The next theorem characterizes the expressibility of source queries in
an OBDA specification. It has several interesting consequences. First,
it implies that the UCQ $\Mbf(q_s)$ is a realization of a UCQ $q_s$
over \Sbf if there is any such realization.  This is well known in the
case without an ontology
\cite{DBLP:journals/tods/NashSV10,DBLP:journals/tcs/Afrati11} and is
implicit in \cite{DBLP:conf/dlog/Cima17} for a rather special case of
OBDA.  Second, the theorem provides a polynomial time reduction of
expressibility to verification: $q_s$ is expressible in \Smc iff
$\Mbf(q_s)$ is a realization of $q_s$ in \Smc. And third, it shows
that if $q_s$ is a CQ, then CQ-expressibility coincides with
UCQ-expressibility. Thus, all lower bounds for CQ-to-CQ expressibility
also apply to (U)CQ-to-UCQ expressibility and all upper bounds for
UCQ-to-UCQ verification and expressibility also apply to the
corresponding CQ-to-(U)CQ case.
\begin{theorem}
\label{thm:maincharact} 
Let $\Smc=(\Omc,\Mbf,\Sbf)$ be an OBDA specification from
$(\mbox{\text{FO}(=)},\text{GAV})$, $q_s$ a UCQ over \Sbf, and $q_r$
an infinitary UCQ rewriting of the OMQ
$Q=(\Omc,\mn{sch}(\Mbf),\Mbf(q_s))$.
Then $q_s$ is UCQ-expressible in \Smc iff $\Mbf^-(q_r)
\subseteq_\Sbf q_s$. Moreover, if this is the case then $\Mbf(q_s)$ is
a realization of $q_s$ in~\Smc.
\end{theorem}
\noindent
\begin{samepage}
\begin{proof}
  We first observe that
  \begin{itemize}
  \item[(a)] if $\Mbf^-(q_r) \subseteq_\Sbf q_s$, then $\Mbf(q_s)$ is
    a realization of $q_s$ in \Smc.
  \end{itemize}
  This actually follows from Theorem~\ref{thm:realization} because
  $q_s \subseteq_\Sbf \Mbf^-(q_r)$ always holds. In fact, since
  $\Mbf(q_s)$ is the actual query in $Q$ and since $q_r$ is a
  rewriting of $Q$, we have $\Mbf(q_s) \subseteq_{\mn{sch}(\Mbf)}
  q_r$; applying Point~3 of Lemma~\ref{lem:patterns} then yields $q_s
  \subseteq_\Sbf \Mbf^-(q_r)$.  \smallskip

  Note that ($a$) establishes the ``if'' part of
  Theorem~\ref{thm:maincharact}.  In view of
  Theorem~\ref{thm:realization} and by ($a$), we can prove both the
  ``only if'' and the ``Moreover'' part by showing that if there is any
  realization $q_t$ of $q_s$ in \Smc, then $\Mbf(q_s)$ is a
  realization of~$q_s$. 

\smallskip 
  Thus assume that $q_t$ is such a realization and let $Q'$ be the OMQ
  $(\Omc,\mn{sch}(\Mbf),q_t)$ and $q'_r$ a UCQ-rewriting of $Q'$. We
  aim to show that
  \begin{itemize}

  \item[(b)] $\Mbf^-(q_r) \subseteq_\Sbf \Mbf^-(q'_r)$.

  \end{itemize}
  This suffices since Theorem~\ref{thm:realization} yields
  $\Mbf^-(q'_r) \subseteq_\Sbf q_s$ and composing (b) with this containment
  gives $\Mbf^-(q_r) \subseteq_\Sbf q_s$ that yields the desired
  result because of (a).

  To establish (b), by Point~2 of Lemma~\ref{lem:patterns} it suffices
  to show $q_r \subseteq_{\mn{sch}(\Mbf)} q'_r$. 
% Let $p$ be a CQ from
%   $q_r$.
  From Theorem~\ref{thm:realization}, we get $q_s \subseteq_{\Sbf}
  \Mbf^-(q'_r)$. Point~3 of Lemma~\ref{lem:patterns} gives $\Mbf(q_s)
  \subseteq_{\mn{sch}(\Mbf)} q'_r$. By the semantics of certain
  answers, this implies $(\Omc,\mn{sch}(\Mbf),\Mbf(q_s))
  \subseteq_{\mn{sch}(\Mbf)} (\Omc,\mn{sch}(\Mbf),q'_r)$. Since the
  former OMQ is just $Q$ and $q_r$ is a rewriting of $Q$, we get $q_r
  \subseteq_{\mn{sch}(\Mbf)} (\Omc,\mn{sch}(\Mbf),q'_r)$. It thus
  remains to show $(\Omc,\mn{sch}(\Mbf),q'_r) \subseteq q'_r$, which
  is exactly the statement of Lemma~\ref{lem:doublebackchain} in the
  appendix.
\end{proof}
\end{samepage}
The following corollary of Theorem~\ref{thm:maincharact} shows that
while making the ontology logically stronger might make some source
queries inexpressible (see Example~\ref{ex:main}), it never results in
additional such queries becoming expressible.
\begin{corollary}
  Let $\Smc_i=(\Omc_i,\Mbf,\Sbf)$ $i \in \{1,2\}$, be OBDA specifications
  from $[\text{FO},\text{GAV}]$ with $\Omc_1 \models \Omc_2$, $\Qmc
  \in \{ \text{CQ}, \text{UCQ} \}$ and $q_s$ from $\Qmc$.  Then
  $\Qmc$-expressibility of $q_s$ in $\Smc_1$ implies
  $\Qmc$-expressibility of $q_s$ in $\Smc_2$.
\end{corollary}
\noindent
\begin{proof}
  Assume that $q_s$ is $\Qmc$-expressible in $\Smc_1$. Then
  Theorem~\ref{thm:maincharact} gives that $\Mbf(q_s)$ is a
  realization, and this query is also from $\Qmc$. We show that
  $\Mbf(q_s)$ is also a realization of $q_s$ in $\Smc_2$.  Let
  $q_{r,i}$ be the canonical infinitary UCQ rewriting of the OMQ $Q_i
  = (\Omc_i,\mn{sch}(\Mbf),\Mbf(q_s))$, $i \in \{1,2\}$. By
  Theorem~\ref{thm:realization}, $q_s \equiv_\Sbf
  \Mbf^-(q_{r,1})$. Since $\Omc_1 \models \Omc_2$, we have $Q_2
  \subseteq_{\mn{sch}(\Mbf)} Q_1$.  This clearly implies that every CQ
  in $q_{r,2}$ is also in $q_{r,1}$.  Thus $q_s \equiv_\Sbf
  \Mbf^-(q_{r,1})$ implies $q_s \supseteq_\Sbf \Mbf^-(q_{r,2})$. It
  remains to argue that $q_s \subseteq_\Sbf \Mbf^-(q_{r,2})$. Since
  $q_{r,2}$ is a rewriting of $Q_2$, we have $Q_2
  \subseteq_{\mn{sch}(\Mbf)} q_{r,2}$. By the semantics and definition
  of $Q_2$, $\Mbf(q_s) \subseteq_{\mn{sch}(\Mbf)} Q_2$ and thus
  $\Mbf(q_s) \subseteq_{\mn{sch}(\Mbf)} q_{r,2}$. Point~3 of
  Lemma~\ref{lem:patterns} yields $q_s \subseteq_{\Sbf}
  \Mbf^-(q_{r,2})$ as desired.
\end{proof}

\section{Expressibility and Verification in DL-Lite}

We consider OBDA specifications in which the ontology is formulated in
a dialect of DL-Lite. The distinguishing feature of logics from this
family is that finite UCQ rewritings of OMQs always exist. Therefore,
Theorems~\ref{thm:realization} and~\ref{thm:maincharact} immediately
imply decidability of the verification and expressibility problem,
respectively. It is, however, well known that UCQ rewritings can
become exponential in size \cite{DBLP:journals/ai/GottlobKKPSZ14} and
thus optimal complexity bounds are not immediate. 

We consider the dialect DL-Lite$^\Rmc_{\mn{horn}}$ as a typical
representative of the DL-Lite family of logics. However, our results
also apply to many other dialects since their proof rests only on the
following properties, established in
\cite{DBLP:journals/jair/ArtaleCKZ09}.
\newcommand{\Lang}{\mathcal{L}}
\begin{samepage}
\begin{theorem} \label{thm:dlliteprops}
  In DL-Lite$^\Rmc_{\mn{horn}}$,
\begin{enumerate}
\item all OMQs $Q$ have a UCQ-rewriting in which all CQs are of size
  polynomial in $|Q|$;
 \item OMQ evaluation is in \NPclass in combined complexity.
\end{enumerate}
\end{theorem}
\end{samepage}
We remark that the results presented in this section are related to
those obtained in \cite{DBLP:conf/dlog/Cima17}, where the
DL-Lite$_{\Amc,id}$ dialect of DL-Lite is considered, mappings are
GLAV, and queries CQs. A main difference is that Cima's technical
results concern rewritings that are complete but not necessarily
sound, which corresponds to replacing
`$\mn{ans}_{q_s}(D)=\mn{cert}_{Q}(\Mbf(D))$' in
Definition~\ref{def:main} with
`$\mn{ans}_{q_s}(D)\subseteq\mn{cert}_{Q}(\Mbf(D))$'. Some of his
technical constructions are similar to ours. Note that
DL-Lite$_{\Amc,id}$ also satisfies the conditions from
Theorem~\ref{thm:dlliteprops} and thus our results apply to
$[\text{DL-Lite}_{\Amc,id},\text{GAV}]$ as well.

%
% The condition of Theorem~\ref{thm:dlliteupper} are satisfied by many
% dialects in the DL-Lite family. {\color{blue} Give examples in the
% literature. Maybe discuss a little of the difficulties that we get
% because we define rewritings relative to all databases and not just
% the consistent ones. Obtain a Corollary for those dialects.}
%
% clu: not true!
% In the proof of Theorem~\ref{thm:dlliteupper} we need the following
% lemma, which connects the two assumptions of the theorem:
%
For an OMQ $Q=(\Omc,\Sbf,q)$, with \Omc formulated in
\mbox{FO$(=)$} and $q$ a UCQ, the \emph{canonical UCQ-rewriting of
  size $n$} is the UCQ $q_c$ that consists of all pairs $(\Amc,\abf)$ viewed
as a CQ where $\abf \in \mn{cert}_Q(\Amc)$ and $|\Amc| \leq n$. The
following lemma is interesting in connection with Point~1 of
Theorem~\ref{thm:dlliteprops} as it allows us to concentrate on
canonical UCQ rewritings of polynomial size.
%
% {\color{blue} Should the following really be a lemma? Maybe it is too
% simple\dots}
\begin{restatable}{lemma}{lemcanonicalrewriting}
\label{lem:canonicalrewriting} Let $Q=(\Omc,\Sbf,q)$
  be an OMQ with \Omc formulated in FO$(=)$ and $q$ a UCQ. If $Q$ has
  a UCQ-rewriting $q_r$ in which all CQs are of size at most $n$, then
  the canonical UCQ-rewriting $q_c$ of size $n$ is also a rewriting of $Q$.
\end{restatable}
We are now ready to establish the upper bound. % Recall that, due
% to the remark before Theorem~\ref{thm:maincharact}, the following
% result also captures the CQ-to-CQ and CQ-to-UCQ cases (which are
% identical). The same is true for all other upper bounds in this paper.
%
\begin{theorem}
  \label{thm:dlliteupper} In
  $[\text{DL-Lite}^\Rmc_{\mn{horn}},\text{GAV}]$, the UCQ-to-UCQ
  expressibility and verification problems are in $\Pi^p_2$.
\end{theorem}
\noindent
\begin{proof} 
  As remarked before Theorem~\ref{thm:maincharact}, expressibility
  polynomially reduces to verification and thus it
  suffices to consider the latter.  Hence let the following be given:
  an OBDA specification $\Smc = (\Omc,\Mbf,\Sbf)$ from
  $[\text{DL-Lite}^\Rmc_{\mn{horn}},\text{GAV}]$, a UCQ $q_s$ over
  schema $\Sbf$, and a UCQ $q_t$ over the schema
  $\mn{sch}(\Mbf)$. % We need
% to provide a $\Pi^p_2$-algorithm that checks whether $q_t$ is a
% realization of $q_s$. 
Let $n$ be the size of this input.

Let $Q = (\Omc,\mn{sch}(\Mbf),q_t)$ and note that the size of $Q$
is polynomial in $n$. By Point~1 of Theorem~\ref{thm:dlliteprops}, we
can assume that $Q$ has a UCQ-rewriting in which all CQs are of
size $P(n)$, $P$ a polynomial. By Lemma~\ref{lem:canonicalrewriting},
we can even assume that this rewriting is the canonical UCQ-rewriting
$q_c$ of size $P(n)$.  By Theorem~\ref{thm:realization}, $q_t$ is thus
a realization of $q_s$ iff $q_s \equiv_\Sbf \Mbf^-(q_c)$. We show that
both inclusions of this equivalence can be checked in $\Pi^p_2$.

First, consider the inclusion $q_s \subseteq_\Sbf \Mbf^-(q_c)$. It
holds iff for every $q$ in $q_s$, there is a $p$ in $\Mbf^-(q_c)$ and
a homomorphism $p \to q$. This condition can be checked even in
\NPclass: iterate over all CQs $q$ in $q_s$ (of which there are at
most $n$), guess a disjunct $p$ from $\Mbf^-(q_c)$, and verify in
\NPclass that $p \to q$.

To guess a $p$ in $\Mbf^-(q_c)$, it suffices to guess a pair
$(\Amc,\abf)$ in $q_c$ and suitable mappings from $\Mbf$ for every fact
in $\Amc$, which determine $p$. Then, $p$ can be computed in polynomial
time from $\Amc$ and these suitable mappings. We guess the pair
$(\Amc,\abf)$ from $q_c$ by guessing an arbitrary ABox $\Amc$ of size at
most $P(n)$ and then verifying that $\abf \in \mn{cert}_Q(\Amc)$. By
Point~2 of Theorem~\ref{thm:dlliteprops} this verification is possible
in $\NPclass$.

% To guess $p$, it suffices to guess a CQ of size bounded by $P'(n)$
% where $P'$ is a polynomial that bounds the size of $\Mbf^-(p)$ for CQs
% of size at most $P(n)$; clearly, such a polynomial exists. It remains
% to verify that $p$ is in $\Mbf^-(q_c)$. This is done as follows: guess
% a pair $(\Amc,\abf)$ of an ABox $\Amc$ and tuple $\abf \in
% \mn{adom}(\Amc)$ such that the size of \Amc is at most $P(n)$.
% Additionally, guess a suitable mapping from $\Mbf$ for every fact in
% $\Amc$. By definition of $\Mbf^-$, it is clear that we can check in
% polynomial time whether $p = \Mbf^-(\Amc,\abf)$. It remains to be
% checked that $(\Amc,\abf)$ is a CQ in $q_c$. By the definition of
% $q_c$, this amounts to verifying $\abf \in \mn{cert}_Q(\Amc)$, which
% by Point~2 of Theorem~\ref{thm:dlliteprops} is possible in $\NPclass$.

We next consider the inclusion $\Mbf^-(q_c) \subseteq_\Sbf q_s$. This
holds iff for every $p$ in $\Mbf^-(q_c)$, there is a CQ $q$ in $q_s$
such that $q \to p$. We can thus universally guess a $p$ in
$\Mbf^-(q_c)$, then iterate over all CQs $q$ in $q_s$, and for each
such $q$ check in \NPclass whether $q \to p$. For universally
guessing $p$, we actually guess a CQ $p$ of size at most $P'(n)$
and then verify that it is in $\Mbf^-(q_c)$. It has already been
argued above that this is possible in $\NPclass$. Overall, we obtain
a $\Pi^p_2$-algorithm, as desired.
\end{proof}
We next show that the expressibility problem in
$[\text{DL-Lite}^\Rmc_{\mn{horn}},\text{GAV}]$ is $\Pi^p_2$-hard, and
thus the same holds for the verification problem. Interestingly, the
lower bound already applies when the ontology is empty and the source
query is a CQ.  As noted in the introduction, this shows that expressibility
of a source CQ as a (U)CQ over UCQ views is $\Pi^p_2$-hard, and in
fact it is $\Pi^p_2$-complete by Theorem~\ref{thm:dlliteupper}. This
corrects a (very likely) erroneous statement of \NPclass-completeness in
\cite{Levy95}.
\begin{restatable}{theorem}{thmpiptwolower}
\label{thm:pip2lower}
The CQ-to-CQ expressibility problem is $\Pi^p_2$-hard for GAV mappings
and the empty ontology. 
\end{restatable}
The proof is by reduction of validity of $\forall \exists$-QBFs. By
Theorem~\ref{thm:maincharact}, expressibility in the absence of an
ontology amounts to checking the containment $\Mbf^-(\Mbf(q_s))
\subseteq q_s$, which is equivalent to the $\forall \exists$-statement
that for all $p \in \Mbf^-(\Mbf(q_s))$ there is a homomorphism $q_s
\to p$. Hence we encode a $\forall \exists$-quantified Boolean formula
such that the outer universal quantifiers correspond to the different
choices of mappings when taking a $p \in \Mbf^-(\Mbf(q_s))$, whereas
the inner existential quantifiers of the formulas correspond to
homomorphisms $q_s \to p$.

\section{Expressibility in \ELHI: Upper Bound for Rooted Queries}

We show that the expressibility problem in $[\ELHI,\text{GAV}]$ is in
\coNExpTime when the source query is a rooted UCQ. Here, a CQ $q$ is
\emph{rooted} or an \emph{rCQ} if every variable from $q$ is reachable
from an answer variable in the hypergraph
$H_q:=(\mn{var}(q),\{\{x_1,\dots,x_n\} \mid R(x_1,\dots,x_n) \in q\})$
and a UCQ is \emph{rooted} or an \emph{rUCQ} if every CQ in it is
rooted.  In practice, many relevant queries are rooted. Our aim is
thus to prove the following result.
\begin{theorem}
\label{thm:ELHIupperrooted}
%Let $\Qmc_s, \Qmc_t \in \{rCQ, rUCQ\}$. 
  In $[\ELHI, \text{GAV}]$, the rUCQ-to-UCQ expressibility problem is
  in $\coNExpTime$.
\end{theorem}
To prepare for lifting the result from expressibility to verification
later, we actually establish a slightly more general result as needed.
Note that the following implies Theorem~\ref{thm:ELHIupperrooted}
since, by Theorem~\ref{thm:maincharact}, we can simply use $\Mbf(q_s)$
for $q_t$. % and complement the result.
\begin{theorem}
\label{thm:ELHIupperrootedgen}
Given an OBDA setting $\Smc=(\Omc,\Mbf,\Sbf)$ from
$[\ELHI, \text{GAV}]$, an rUCQ $q_s$ over \Sbf, and a UCQ $q_t$ over
$\mn{sch}(\Mbf)$, it is in \coNExpTime to decide whether
$\Mbf^-(q_r) \subseteq_\Sbf q_s$, where $q_r$ is an infinitary UCQ-rewriting
of the OMQ $Q=(\Omc,\mn{sch}(\Mbf),q_t)$.
\end{theorem}
To prove Theorem~\ref{thm:ELHIupperrootedgen}, we now describe a
\NExpTime algorithm for deciding the complement of the problem
described there: we want to check whether $\Mbf^-(q_r) \not\subseteq
q_s$, that is, whether there is a CQ $p$ in the UCQ $\Mbf^-(q_r)$ such
that $q \not\rightarrow p$ for all CQs $q$ in $q_s$. Because
all rewritings of $Q$ are equivalent, it suffices to prove the theorem
for any particular infinitary UCQ-rewriting $q_r$ of $Q$. We choose to
work with the canonical one introduced at the beginning of the
characterizations section. The algorithm is as follows:
\begin{enumerate}
% \item Compute $\Mbf(q_s)$ in polynomial time.

\item Guess a $\mn{sch}(\Mbf)$-ABox $\Amc$ such that
  $|\mn{adom}(\Amc)| \leq |q_t| + |q_t| \cdot |\Omc|^{|q_s|+1}$ % , where
  % $n = \max(|q_s|,|q_t|) + 1$,
  and a tuple $\abf$ in $\Amc$ of the same arity as~$q_t$.
% , where 
%   $n = \max(|q_s|,|\Mbf(q_s)|) + 1$;
% Guess an pseudo tree-shaped ABox $\Amc$ with a core no larger than
% $|M(q)|$ and with trees of branching degree not larger than $|\Omc|$
% and depth not larger than $n$. Note that the $|\Amc|$ is exponential
% in the size of the input. Note that the size $\Amc$ is exponential in
% the size of $q$ and polynomial in the size $\Omc$.
% Additionally, guess a tuple $\abf$ in the core of $\Amc$ of same length
% as $\xbf$.
\label{guess abox}

\item Verify that $\abf \in \mn{cert}_Q(\Amc)$ to make sure that
  $(\Amc,\abf)$ viewed as a CQ is in the UCQ $q_r$. This can be done
  by an algorithm that is exponential in $|\Omc|$ and $|q_t|$, but
  only polynomial in $|\Amc|$; see for
  example~\cite{DBLP:conf/lpar/KrisnadhiL07}. Hence, the overall
  running time is single exponential in the size of the original
  input.
\label{verify truth}

\item Guess a disjunct $p$ from the UCQ $\Mbf^-(\Amc,\abf)$ by
guessing, for each fact $\alpha$ in \Amc, a suitable mapping from
$\Mbf$. Note that both $\Amc$ and $p$ are of single exponential size.
\label{choose disjunct}

\item Verify that $q \not\rightarrow p$ for all CQs $q$ in the rUCQ
$q_s$. This can be done in single exponential time using brute force.

\label{verify no hom}
\end{enumerate}
This is clearly a \NExpTime algorithm. %It remains to show correctness.
\begin{restatable}{lemma}{lemnexpalgcorr}
\label{lem:nexpalgcorr}
  The algorithm decides the complement of the problem in
  Theorem~\ref{thm:ELHIupperrootedgen}.
\end{restatable}
It is easy to verify the soundness part: a successful run of the
algorithm identifies $p$ as a CQ in $\Mbf^-(q_r)$ such that for any
disjunct $q$ of $q_s$, $q \not\rightarrow p$. Completeness is less
obvious, mainly because of the magical size bound used in Step~1. We
need some preliminaries.

% Maybe I should use the very same definition here as in the other
% papers? There tree shaped is defined with reference to the
% undirected graph underlying the ABox.
An ABox $\Amc$ is \emph{tree-shaped} if the undirected graph $G_\Amc =
(\mn{adom}(\Amc),\{\{a,b\} \mid r(a,b) \in \Amc\})$ is acyclic,
connected and $r(a,b) \in \Amc$ implies that $s(a,b) \notin \Amc$ for all
role names $s \neq r$ and that $s(b,a) \notin \Amc$ for all role names $s$.
An ABox $\Amc$ is \emph{pseudo tree-shaped with core} $\Cmc \subseteq
\Amc$ if there is a tree-shaped ABox $\Amc_a$ with root $a$ for every $a \in \mn{adom}(\Cmc)$, with mutually
disjoint domains, such that $\Amc = \Cmc \cup \bigcup_{a \in
\mn{adom}(\Cmc)} \Amc_a$.
%
% $\Amc$ if there is some enumeration $\{c_1,\dots,c_n\}$ of the
% individuals in $\Cmc$ and for each $i \in \{1,\dots,n\}$ there is a
% tree-shaped ABox $\Amc_i$ with root $c_i$ such that the individuals
% $\Amc_i$ and $\Amc_j$ are mutually disjoint for $i \neq j$ and
% $\Amc$ is the union of $\Cmc$ and all the $\Amc_i$ for $i \in
% \{1,\dots,n\}$.
The \emph{outdegree} of $\Amc$ is the maximal outdegree of the trees
underlying any of the $\Amc_a$.

The following lemma is an adaptation of Proposition~23 in Appendix~B
of \cite{BHLW-IJCAI16}:
\begin{lemma} \label{lem:unravel} Consider an \ELHI-ontology $\Omc$,
an OMQ $Q = (\Omc,\Sbf,q)$ with $q$ a UCQ, an \Sbf-ABox $\Amc$, and an
answer $\abf \in \mn{cert}_Q(\Amc)$. Then there is a pseudo
tree-shaped \Sbf-ABox $\Amc'$ and a tuple $\abf'$ in the
core of $\Amc'$ such that \begin{enumerate} \item the core of
$\Amc'$ is not larger than $|q|$ and the outdegree of \Amc is not
larger than $|\Omc|$; \item $\abf' \in \mn{cert}_Q(\Amc')$;
 % \item $\abf' \in \mn{cert}_Q(\Cmc' \cup \{ A(a) \mid \Amc',\Omc
 % \models A(a),\; a \in \mn{dom}(\Cmc')\})$;
 \item there is a homomorphism $h$ from $\Amc'$ to $\Amc$ with
$h(\abf')=\abf$. \end{enumerate} \end{lemma}
% \begin{lemma} \label{lem:unravel} Consider an \ELI-ontology $\Omc$,
% an OMQ $Q = (\Omc,\Sbf,q)$ with $q$ a UCQ, an \Sbf-ABox $\Amc$, and
% an answer $\abf \in \mn{cert}_Q(\Amc)$. Then there is a pseudo
% tree-shaped \Sbf-ABox $\Amc'$ and a tuple $\abf'$ in the core of
% $\Amc'$ such that \begin{enumerate} \item the core of $\Amc$ is not
% larger than $|q|$ and the outdegree of \Amc is not larger than
% $|\Omc|$; \item $\abf' \in \mn{cert}_Q(\Amc')$; \item there is a
% homomorphism $h$ from $\Amc'$ to $\Amc$ with $h(\abf')=\abf$.
% \end{enumerate} \end{lemma}
%
%
% This remark might not be needed for later. But maybe we should
% actually define a terminology or notation for the rewriting because
% we want to use it later??
%
We are now ready for the completeness part of
Lemma~\ref{lem:nexpalgcorr}. % The complete argument is in the appendix, we
% just give a sketch here.
Assume that there is a CQ $p$ in
$\Mbf^-(q_r)$ such that for any disjunct $q$ of $q_s$, $q
\not\rightarrow p$. Since $q_r$ is the canonical infinitary
UCQ-rewriting of $Q$, $p$ is of the form $\Mbf^-(\Amc,\abf)$ where
$\Amc$ is a $\mn{sch}(\Mbf)$-ABox with $\abf \in \mn{cert}_Q(\Amc)$.
Let $\Amc'$ be the pseudo tree-shaped ABox and $\abf'$ the answer
whose existence is guaranteed by Lemma~\ref{lem:unravel}. Moreover,
let $\Amc''$ be the result of removing from $\Amc'$ all facts
that contain at least one constant whose distance from the core is
larger than $|q_s|$ and adding for all constants $a$ of
distance exactly $|q_s|$ from the core both $A(a)$ for all concept names
$A$ in $\mn{sch}(\Mbf)$ and $r(a,a)$ for all role names $r$ in
$\mn{sch}(\Mbf)$. We show in the appendix that $\abf' \in
\mn{cert}_Q(\Amc'')$ and that there is a CQ $p$ in
$\Mbf^-(\Amc'',\abf')$ such that for any disjunct $q$ of $q_s$, $q
\not\rightarrow p$ (this depends on $q_s$ being rooted). The former
implies that $(\Amc'',\abf')$, seen as CQ, is a disjunct in $q_r$ and
hence Step~2 of the algorithm succeeds. The latter guarantees that
Step~4 succeeds. Moreover, $\Amc''$ satisfies the size bound given in
Step~1 of the algorithm.

\section{Expressibility in \ELHI: Upper Bound for Unrestricted Queries}

We consider the UCQ-to-UCQ expressibility problem in
$[\ELHI,\text{GAV}]$, that is, we drop the assumption from the
previous section that the source query is rooted. This increases the
complexity from \coNExpTime to 2\ExpTime. Note that similar effects
have been observed in the context of different reasoning problems in
\cite{DBLP:conf/cade/Lutz08,BHLW-IJCAI16}. In this section, we show
the upper bound.
\begin{theorem}
\label{thm:ELHIuppernotrooted}
%Let $\Qmc_s, \Qmc_t \in \{rCQ, rUCQ\}$. 
  In $[\ELHI, \text{GAV}]$, the UCQ-to-UCQ expressibility problem is
  in $2\ExpTime$.
\end{theorem}
As in the rooted case, we again prove a slightly more general result
that can be reused when studying the verification problem. We can
obtain Theorem~\ref{thm:ELHIuppernotrooted} from the following 
by setting $q_t = \Mbf(q_s)$ and applying
Theorem~\ref{thm:maincharact}.
\begin{theorem}
\label{thm:ELHIuppernotrootedgen}
Given an OBDA setting $\Smc=(\Omc,\Mbf,\Sbf)$ from
$[\ELHI, \text{GAV}]$ a UCQ $q_s$ over \Sbf, and a UCQ $q_t$ over
$\mn{sch}(\Mbf)$, it is in 2\ExpTime to decide whether
$\Mbf^-(q_r) \subseteq_\Sbf q_s$, where $q_r$ is an infinitary UCQ-rewriting
of the OMQ $Q=(\Omc,\mn{sch}(\Mbf),q_t)$.
\end{theorem}
To prove Theorem~\ref {thm:ELHIuppernotrootedgen}, we start by
choosing a suitable UCQ-rewriting $q_r$. Instead of working with the
canonical infinitary UCQ-rewriting, here we prefer to use the UCQ that
consists of all pairs $(\Amc,\abf)$ viewed as a CQ and where $\Amc$ is
a pseudo tree-shaped $\mn{sch}(\Mbf)$-ABox that satisfies $\abf \in
\mn{cert}_Q(\Amc)$ and is of the dimensions stated in
Lemma~\ref{lem:unravel}, that is, the core of $\Amc$ is not larger
than $|q|$ and the outdegree of \Amc is not larger than $|\Omc|$. Due
to that lemma, $q_r$ clearly is an infinitary UCQ-rewriting of $Q$.

We give a decision procedure for the complement of the problem in
Theorem~\ref {thm:ELHIuppernotrootedgen}. We thus have to decide
whether there is a pseudo-tree shaped $\mn{sch}(\Mbf)$-ABox \Amc of
the mentioned dimensions, an $\abf \in \mn{cert}_Q(\Amc)$, and a CQ
$p$ in the UCQ $\Mbf^-(\Amc,a)$ such that $q \not\rightarrow p$ for
all CQs $q$ in $q_s$. This can be done by constructing a two-way
alternating parity tree automaton (TWAPA) \Amf on finite trees that
accepts exactly those trees that represent a triple $(\Amc,\abf,p)$
with the components as described above, and then testing whether the
language accepted by \Amf is empty.

In the following, we detail this construction. We reuse some encodings
and notation from a TWAPA construction that is employed in
\cite{BHLW-IJCAI16} to decide OMQ containment as this saves us from
redoing certain routine work.  A \emph{tree} is a non-empty (and
potentially infinite) set $T \subseteq \mathbb{N}^*$ closed under
prefixes. We say that $T$ is \emph{$m$-ary} if $T \subseteq
\{1,\dots,m\}^*$ and call the elements of $T$ the \emph{nodes} of the
tree and $\varepsilon$ its \emph{root}. For an alphabet $\Gamma$, a
\emph{$\Gamma$-labeled tree} is a pair $(T,L)$ with $T$ a tree and
$L:T \rightarrow \Gamma$ a node labeling function.

We encode triples $(\Amc,\abf,p)$ as finite $(|\Omc|\cdot |q_t|)$-ary
$\Sigma_\varepsilon \cup \Sigma_N$-labeled trees, where
$\Sigma_\varepsilon$ is the alphabet used for labeling the root node
and $\Sigma_N$ is for non-root nodes. These alphabets are different
because the root of a tree represents the core part of a pseudo
tree-shaped ABox whereas each non-root node represents a single
constant of the ABox that is outside the core. Let $\Csf_{\mn{core}}$ 
be a fixed set of $|q_t|$ constants.
%
% TODO: define what is meant by a suitable mapping in the preliminaries.
%
Formally, the alphabet $\Sigma_\varepsilon$ is the set of all triples
$(\Bmc,\abf,\mu)$ where \Bmc is a $\mn{sch}(\Mbf)$-ABox of size at
most $|q_t|$ that uses only constants from $\Csf_{\mn{core}}$, \abf is
a tuple over $\Csf_{\mn{core}}$ whose length matches the
arity of $q_s$, and $\mu$ associates every fact $\alpha$ in \Bmc with
a mapping $\mu(\alpha) \in \Mbf$ that is suitable for $\alpha$.  The
alphabet $\Sigma_N$ consists of all triples $(\Theta,M,\mu)$ where
$\Theta \subseteq (\NC \cap \mn{sch}(\Mbf)) \uplus \{ r, r^{-} \mid r
\in \NR \cap \mn{sch}(\Mbf)\} \uplus \Csf_{\mn{core}}$ contains
exactly one (potentially inverse) role and at most one element of
$\Csf_{\mn{core}}$, $M \in \Mbf$ is a mapping suitable for the fact
$r(a,b)$ with $r$ the unique role name in $\Theta$, and $\mu$ assigns
to each $A \in \Theta$ a mapping $\mu(A) \in \Mbf$ suitable for the
fact $A(a)$.\footnote{Here, $a$ and $b$ are arbitrary but fixed
  constants.} In the following, a \emph{labeled tree} generally means
a $(|\Omc| \cdot |q_t|)$-ary $\Sigma_\varepsilon \cup
\Sigma_N$-labeled tree.

A labeled tree is \emph{proper} if (i)~the root node is labeled with a
symbol from~$\Sigma_\varepsilon$, (ii)~each child of the root is
labeled with a symbol from $\Sigma_N$ that contains an element of
$\Csf_{\mn{core}}$, (iii)~every other non-root node is labeled with a
symbol from $\Sigma_N$ that contains no constant name, and
(iv)~every non-root node has at most $|\Omc|$ successors and (v)~for
every $a \in \Csf_{\mn{core}}$, the root node has at most $|\Omc|$
successors whose label includes $a$.  A proper labeled tree $(T,L)$
with $L(\varepsilon)=(\Bmc,\abf,\mu)$ \emph{encodes} the triple
$(\Amc,\abf,p)$ where $\Amc$ is the
ABox % whose constants are $\mn{adom}(\Bmc)$ plus all non-root nodes of
% $T$ viewed as constants, and whose assertions are
\[
\begin{array}{l}
   \Bmc \cup \{ A(x) \mid A \in \Theta(x) \}\\[0.5mm] % \cap \NC  \}\\
   \cup\; \{r(b,x) \mid \{b,r\} \subseteq \Theta(x)\} \cup \{r(x,b) \mid 
  \{b,r^{-}\} \subseteq \Theta(x)\}\\[0.5mm]
   \cup\; \{ r(x,y) \mid r \in \Theta(y), y \text{ is a child of } x, 
    \Theta(x) \in \Sigma_N\}\\[0.5mm]  %last item ensure x not the root
   \cup\; \{ r(y,x) \mid  r^{-} \in \Theta(y), y \text{ is a child of } x, 
  \Theta(x) \in \Sigma_N \}, % same here, x not root
\end{array}
\]
 % \item $\abf$ is the tuple in the label of the root node.
$\Theta(x)$ denoting $\Theta$ when $L(x)=(\Theta,M,\mu)$ (and
undefined otherwise), and where $p$ is the CQ from $\Mbf^-(\Amc)$ that
can be obtained by choosing for every fact in $\Amc$ the suitable
mapping from \Mbf assigned to it by $L$.

The desired TWAPA $\Amf$ is obtained as the intersection of two TWAPAs
$\Amf_1$ and $\overline{\Amf}_2$, where $\Amf_1$ accepts exactly the
proper labeled trees $(T,L)$ that encode a pair $(\Amc,\abf,p)$ with
$\abf \in \mn{cert}_Q(\Amc)$ and $\overline{\Amf}_2$ is obtained as
the complement of an automaton $\Amf_2$ that accepts a proper labeled
tree $(T,L)$ encoding a pair $(\Amc,\abf,p)$ iff $q \to p$ for some CQ
$q$ in $q_s$.  In fact, the automaton $\Amf_1$ is what we can reuse
from \cite{BHLW-IJCAI16}, see Point~1 in Proposition~13 there. The
only difference is that our trees are decorated in a richer way, so
in our case the TWAPA ignores the part of the labeling that is
concerned with mappings from \Mbf. The number of states of $\Amf_1$ is
single exponential in $|q_t|$ and $|\Omc|$.

We now sketch the construction of the automaton $\Amf_2$ for a single
CQ $q$ of $q_s$ (the general case can be dealt with using union). Let
$q_1,\dots,q_k$ be the maximal connected components of $q$. We define
automata $\Amf_{2,1},\dots,\Amf_{2,k}$ where $\Amf_{2,i}$ accepts
$(T,L)$ encoding $(\Amc,\abf,p)$ iff $q_i \to p$, and then intersect
to obtain $\Amf_2$. Let $(T,L)$ be a proper labeled tree. A set $T'
\subseteq T$ is a \emph{subtree} of $T$ if for any $s,t \in T'$, all
nodes from $T$ that are on the shortest (undirected) path from $s$ to
$t$ are in~$T'$.  We use $(T',L)$ to denote the restriction of $(T,L)$
to $T'$ and $\Mbf^-(T',L)$ to denote the subquery of $p$ which
contains only the atoms in $p$ that can be derived from the part of
\Amc generated by the subtree $(T',L)$ of $(T,L)$.

To define $\Amf_{2,i}$, let $\mathcal{C}$ denote the set of all
labeled trees $(T',L)$ of size at most $|q_i|$ such that there is a
proper labeled tree $(T,L)$ and $q_i \to \Mbf^-(T',L)$ with a
homomorphism that only needs to respect the answer variables from $q$
that actually occur in $q_i$ (if there are any, then $T'$ must thus
contain the root of $T$). The automaton $\Amf_{2,i}$ is then
constructed such that it accepts a proper labeled tree $(T,L)$ iff it
contains a subtree from $\mathcal{C}$. It should be clear that such an
automaton can be constructed using only single exponentially many
states. % One can also see that the number of
% states of $\Amf_{2,i}$ is single exponential in the input size because
% $|\mathcal{C}_i|$ is single exponential in the input size. The number
% of unlabelled, rooted trees with at most $|q_i|$ many individuals is
% single exponential in $|q_i|$. To obtain $\mathcal{C}_i$, we label
% such a tree with an alphabet of single exponential size, which yields
% single exponentially many labellings of one such tree.
%
Moreover, it can be verified that $\Amc_{2,i}$ accepts exactly the
desired trees. We obtain an overall automaton with single
exponentially many states which together with the \ExpTime-complete
emptiness problem of TWAPAs gives
Theorem~\ref{thm:ELHIuppernotrootedgen}.

% We argue that $\Amf_{2,i}$ accepts precisely the proper trees $(T,L)$
% encoding a pair $(\Amc,\abf,p)$ such that $q_i \to p$. If $(T,L)$ is accepted
% then it contains a subtee $(T',L)$ from $\mathcal{C}_i$. Hence, $q_i$ maps
% homorphically into $\Mbf^-(T',L)$, which is a subquery of $p$, and
% therefore $q_i \to p$. For the other direction let $q_i \to p$. Because $q_i$
% is connected, there is connected subquery $p'$ of $p$ with size at most
% $|q_i|$ such that $q_i \to p'$. Because $p \in \Mbf^-(\Amc)$ there is then a
% connected subset $\Amc'$ of $\Amc$ mentioning at most $|q_i|$ individuals
% such that $q_i \to \Mbf^-(\Amc')$. Consider then the connected subtree
% $(T',L)$ of $(T,L)$ containing all nodes that encode facts of $\Amc'$.
% This subtree $(T',L)$ is in $\mathcal{C}_i$ because $q_i \to \Mbf^-(T',L)$.
% Hence, $\Amf_{2,i}$ accepts $(T',L)$.

\section{Verification in \ELHI: Upper Bounds}

We show that in $[\ELHI,\text{GAV}]$, the complexity of the
verification problem is not higher than the complexity of the
expressivity problem both in the rooted and in the general case. 
\begin{theorem}
\label{thm:ELHIverfication}
In $[\ELHI, \text{GAV}]$, 
\begin{enumerate}
\item 
the rUCQ-to-UCQ verification problem can be decided in \coNExpTime. 
\item 
the UCQ-to-UCQ verification problem can be decided in 2\ExpTime. 
\end{enumerate}
\end{theorem}
Recall the characterization of realizations from
Theorem~\ref{thm:realization}: a UCQ $q_t$ is a realization of $q_s$
iff $q_s \equiv \Mbf^-(q_r)$, where $q_r$ is a rewriting of
the OMQ $(\Omc,\mn{sch}(\Mbf),q_t)$. The inclusion $q_s \supseteq
\Mbf^-(q_r)$ is already treated by
Theorems~\ref{thm:ELHIupperrootedgen}
and~\ref{thm:ELHIuppernotrootedgen} and thus it remains to show
that the converse inclusion can be decided in the relevant complexity
class. We show that it is actually in \ExpTime even in the unrooted
case. We thus aim to prove the following.
\begin{theorem}
\label{thm:ELHIupperverificationinclusion}
 Given an OBDA setting $\Smc=(\Omc,\Mbf,\Sbf)$ from $[\ELHI,
\text{GAV}]$, a UCQ $q_s$ over \Sbf, and a UCQ $q_t$ over
$\mn{sch}(\Mbf)$, it is in \ExpTime to decide whether $q_s
\subseteq_\Sbf \Mbf^-(q_r)$, where $q_r$ is an infinitary
UCQ-rewriting of the OMQ $Q=(\Omc,\mn{sch}(\Mbf),q_t)$.
\end{theorem}
For what follows, it is convenient to assume that the ontology \Omc is
in \emph{normal form}, that is, all CIs in it are of one of the forms
$\top \sqsubseteq A$, $A_1 \sqcap \cdots \sqcap
A_n \sqsubseteq B$, $A \sqsubseteq \exists r .B$, and $\exists r .A
\sqsubseteq B$ where $A,B$ and all $A_i$ range over concept names and
$r$ ranges over roles. It is well-known that every \ELHI-ontology \Omc
can be converted into an \ELHI-ontology $\Omc'$ in normal form in
linear time such that $\Omc'$ is a conservative extension of $\Omc$ in
the model-theoretic sense \cite{DBLP:books/daglib/0041477}. It is easy
to verify that for the verification problem, we can w.l.o.g.\ assume
the involved ontology to be in normal form.

We again start by choosing a suitable concrete infinitary
UCQ-rewriting to use for $q_r$. As in the previous section, we would
like to use CQs derived from pseudo tree-shaped ABoxes of certain
dimension that entail an answer to the OMQ $Q$, as sanctioned by
Lemma~\ref{lem:unravel}. Here, however, we use a slight strengthening
of that lemma where Condition~2 is replaced with the following
strictly stronger Condition~2$'$, where $\Cmc'$ is the core of the
pseudo tree-shaped ABox $\Amc'$:
$$
\abf' \in \mn{cert}_Q(\Cmc' \cup \{ A(a) \mid \Amc',\Omc
 \models A(a),\; a \in \mn{dom}(\Cmc')\}).
$$
This condition essentially says that, in the universal model of
$\Amc'$ and \Omc (defined in the appendix), there is a homomorphism
$h$ from a CQ in the UCQ $q$ in $Q$ that only involves constants from
the core and `anonymous subtrees' (generated by existential
quantifiers) below them. This is true only since \Omc is assumed to be
in normal form and it is a consequence of the proof of
Lemma~\ref{lem:unravel} where one chooses a homomorphism $h$ from a CQ
in $q$ to the universal model of \Amc and~\Omc, selecting as the core
$\Cmc'$ of the pseudo-tree ABox~$\Amc'$ to be constructed all
constants $a$ from \Amc that are in the range of $h$ or which root an
anonymous subtree that contains an element in the range of $h$, and
then unraveling the rest of \Amc. Summing up, we thus use for $q_r$
the set of all pairs $(\Amc,\abf)$ viewed as a CQ where $\Amc$ is a
pseudo tree-shaped $\mn{sch}(\Mbf)$-ABox with core \Cmc that satisfies
$$
\abf \in \mn{cert}_Q(\Cmc \cup \{ A(a) \mid \Amc,\Omc
 \models A(a),\; a \in \mn{dom}(\Cmc)\})
$$
and is of the dimensions stated in Lemma~\ref{lem:unravel}. 

For deciding $q_s \subseteq_\Sbf \Mbf^-(q_r)$, we need to show that
for every disjunct $q$ in $q_s$ there is a disjunct $p$ in
$\Mbf^-(q_r)$ such that $p \to q$. We can do this for every disjunct
$q$ of $q_s$ separately. Hence let $q$ be such a disjunct.  To find a
CQ $p$ in $\Mbf^-(q_r)$ with $p \to q$, we again aim to utilize
TWAPAs. As in the previous section, let $\Csf_{\mn{core}}$ be a fixed
set of $|q_t|$ constants.  A \emph{homomorphism pattern} for $q_t$ is
a function $\lambda$ that maps every variable $y$ in $q_t$ to a pair
$(a,o) \in \Csf_{\mn{core}} \times
\{\mathsf{core},\mathsf{subtree}\}$. Informally, $\lambda$ is an
abstract description of a homomorphism $h$ from $q_t$ to the universal
model of a pseudo-tree ABox \Amc and \Omc (assume that the core of
\Amc uses only constants from $\Csf_{\mn{core}}$) such
that $h(x)=a$ when $\lambda(x)=(a,\mathsf{core})$ and $h(x)$ is an
element in the anonymous subtree below $a$ when
$\lambda(x)=(a,\mathsf{subtree})$. 

We build one TWAPA $\Amf^\lambda$ for every homomorphism pattern
$\lambda$ (there are single exponentially many). These TWAPAs again
run on $(|\Omc|\cdot |q_t|)$-ary $\Sigma_\varepsilon \cup
\Sigma_N$-labeled trees that encode a triple $(\Amc,\abf,p)$, defined
exactly as in the previous section and from now on are only referred to as
labeled trees. We also use the same notion of properness as in the
previous section. The TWAPA $\Amf^\lambda$ will accept exactly those
labeled trees $(T,L)$ that are proper and encode a triple
$(\Amc,\abf,p)$ such that 
\begin{enumerate}
\item there is a homomorphism $h$ from $q_t(\xbf)$ to the universal
  model of $\Amc$ and $\Omc$ that satisfies $h(\xbf)=\abf$ and follows the
  homomorphism pattern $\lambda$ and
\item $p \to q$.
\end{enumerate}
Note that it would be sufficient to demand in Point~1 that $\abf \in
\mn{cert}_Q(\Amc)$ and recall that, in the previous section, we have
reused an automaton from \cite{BHLW-IJCAI16} which checks exactly this
condition. That automaton, however, has exponentially many states
because it is built using a construction known under various names
such as query splitting, forest decomposition, and squid
decomposition. To attain an \ExpTime upper bound, though, the
automaton $\Amf^\lambda$ can only have polynomially many states.
This is in fact the reason why we have the strengthened Condition~2 of
Lemma~\ref{lem:unravel}.

We construct the automaton $\Amf^\lambda $ as the intersection of three
automata $\Amf_\mn{proper}$, $\Amf^\lambda_1$, and $\Amf_2$, where
$\Amf_\mn{proper}$ accepts if the input tree is proper, $\Amf_1^\lambda$
accepts trees that encode a triple $(\Amc,\abf,p)$ that satisfy
Condition~1 for $\lambda$ and  $\Amf_2$ accepts trees that encode a
triple $(\Amc,\abf,p)$ that satisfy Condition~2. It is easy to build the
automaton $\Amf_\mn{proper}$ and we leave out the details.  The precise
construction of $\Amf_1^\lambda$ and $\Amf_2$ can be found in the
appendix, we only give a brief description here.  Automaton
$\Amf_1^\lambda$ works as follows: it accepts labeled trees $(T,L)$
whose root node label $(\Bmc,\abf,\mu)$ is such that the extension of
the ABox \Bmc with certain facts of the form $A(a)$ results in a
universal model which admits a homomorphism following pattern $\lambda$,
and it then verifies that the facts $A(a)$ used in the extension can be
derived from the ABox encoded by $(T,L)$. The automaton $\Amf_2$ checks
Condition~2 by traversing the input tree once from the root to the
leaves and guessing the homomorphism from $p$ to $q$ along the way. Some
care is required since $p$ is represented only implicitly in the input.

Overall, we obtain single exponentially many automata with
polynomially many states each and we answer `yes' if any of
the automata recognizes a non-empty language. This gives
Theorem~\ref{thm:ELHIupperverificationinclusion}.

\section{Expressibility and Verification in \EL: \\Lower Bounds}

We establish lower bounds that match the upper bounds obtained in the
previous three sections and show that they even apply to
$[\EL,\text{GAV}]$, that is, inverse roles are not required.
\begin{theorem}\label{thm:hardness}~\\[-4mm]
  \begin{enumerate}

  \item In $[\EL,\text{GAV}]$, the rUCQ-to-UCQ expressibility and
    verification problems
    are \coNExpTime-hard.

  \item In $[\EL,\text{GAV}]$, the UCQ-to-UCQ expressibility and
    verification problems are 2\ExpTime-hard

  \end{enumerate}
\end{theorem}
By Theorem~\ref{thm:maincharact}, it suffices to establish the lower
bounds for the expressibility problem.  We prove both points of
Theorem~\ref{thm:hardness} by a reduction from certain OMQ containment
problems. For Point~1, we reduce from the following problem.%  ~\cite{BHLW-IJCAI16}, starting with Point~1. OMQ containment
% in \ELHI is studied in  In the
% containment problems studied in~\cite{BHLW-IJCAI16}, $\Omc_1=\Omc_2$
% is formulated in \ELHI, $q_1$ is an atomic query (AQ) of the form
% $A(x)$, and $q_2$ is a CQ. More precisely, we can assume the
% following.
%
\begin{restatable}{theorem}{nexpbasic}\cite{BHLW-IJCAI16}
\label{lem:nexpbasic} 
Containment between OMQs $Q_1=(\Omc,\Sigma,q_1)$ and
$Q_2=(\Omc,\Sigma,q_2)$ with $\Omc$ an $\ELI$-ontology, $q_1$ an AQ,
and $q_2$ a rooted UCQ is \mbox{\coNExpTime-hard} even when
\begin{enumerate}

\item $q_2(x)$ uses only symbols from $\Sigma$ and

\item no symbol from $\Sigma$ occurs on the right-hand side of a CI in~\Omc.

% \item all occurrences of $\bot$ in \Omc are of the form $C \sqsubseteq \bot$
%   where $C$ is an \ELI-concept in signature $\Sigma$.

\end{enumerate}
\end{restatable}
We first establish Point~1 of Theorem~\ref{thm:hardness} for
$[\ELI,\text{GAV}]$ instead of for $[\EL,\text{GAV}]$ and in a second
step show how to get rid of inverse roles. To reduce the containment
problem in Theorem~\ref{lem:nexpbasic} to rUCQ-to-UCQ expressibility
in $[\ELI,\text{GAV}]$, let $Q_1=(\Omc,\Sigma,A_0(x))$ and
$Q_2=(\Omc,\Sigma,q)$ be as in that theorem.  We define an
OBDA-specification $\Smc = (\Omc',\Mbf,\Sbf)$ and a query $q_s$ over
\Sbf as follows. Let $B$ be a concept name that does not occur in
$Q_1$ and $Q_2$. Set
$$
\begin{array}{rcl}
  \Omc' &=& \Omc \cup \{ A_0 \sqsubseteq B \} \\[1mm]
  \Sbf &=& \Sigma \cup \{ B \}\\[1mm] 
  q_s(x) &=& B(x) \vee q_2(x)
\end{array}
$$
Note that $q_s$ is a rooted UCQ, as required.
Moreover, the set $\Mbf$ of mappings contains $A(x) \rightarrow A(x)$
for all concept names $A \in \Sbf$ and $r(x,y) \rightarrow r(x,y)$
for all role names $r \in \Sbf$. %  ({\color{blue}$M$ is introduced
  % only to ensure that the variable $x$ occurs in every CQ of the UCQ
  % $q_S(x)$; if we do not require that, we can as well drop $M$}).
Informally, the CI $A_0 \sqsubseteq B$ `pollutes' $B$, potentially
preventing the disjunct $B(x)$ of $q_s$ to be expressible, but this
is not a problem if (and only if) $Q_1 \subseteq Q_2$.
\begin{lemma}
\label{lem:nexpCorrMain}
  $Q_1 \subseteq Q_2$ iff $q_s$ is UCQ-expressible in
  \Smc.
\end{lemma}
In short, $Q_1 \not\subseteq Q_2$ iff there is a tree-shaped
$\Sigma$-ABox witnessing this iff such an ABox, viewed as a CQ, is a
disjunct of an infinitary UCQ-rewriting $q_r$ of the OMQ
$Q=(\Omc',\Sbf,q_s)$ iff $q_r \not\subseteq_\Sbf q_s$. The latter is the case
iff $q_s$ is not UCQ-expressible in \Smc by Theorem~\ref{thm:maincharact}
and since $\Mbf(q_s)=q_s$ and $\Mbf^-(q_r)=q_r$.

In the appendix, we describe how to replace the \ELI-ontology \Omc
with an \EL-ontology. The crucial observation is that the hardness
proof from \cite{BHLW-IJCAI16} uses only a single symmetric role $S$
implemented as a composition $r_0^-;r_0$ with $r_0$ a normal role
name, and that it is possible to replace this composition with a
normal role name $r$ in \Omc when reintroducing it in $\Mbf^-(q_r)$
via mappings $r_0(x,y) \wedge r_0(x,z) \rightarrow r(y,z)$ where $q_r$
is an infinitary UCQ-rewriting of the OMQ $Q$ mentioned above.

The 2\ExpTime lower bound in Point~2 of Theorem~\ref{thm:hardness} is
proved similarly, using 2\ExpTime-hardness of a different
containment problem also studied in \cite{BHLW-IJCAI16}.

\section{Conclusion}

We believe that several interesting questions remain. For example, our
lower bounds only apply when the source query is a UCQ and it would be
interesting to see whether the complexity drops when source queries
are CQs. It would also be interesting to consider ontologies
formulated in more expressive DLs such as \ALC. As a first observation
in this direction, we note the following undecidability result, where
$\mathcal{ALCF}$ is \ALC extended with (globally) functional roles. It
is proved by a reduction from the emptiness of AQs w.r.t.\
\ALCF-ontologies \cite{DBLP:journals/jair/BaaderBL16}.
\begin{restatable}{theorem}{thmalcfundec}
\label{thm:alcfundec} 
  In $[\ALCF,\text{GAV}]$, the AQ-to-\Qmc expressibility and verification
  problems are undecidable for any $\Qmc \in \{ \text{AQ}, \text{CQ},
  \text{UCQ} \}$. 
\end{restatable}
Regarding the expressibility problem, we note that the realization
$\Mbf(q_s)$ identified by Theorem~\ref{thm:maincharact} does not use
any symbols introduced by the ontology and, in fact, is also a
realization regarding the empty ontology. It would be interesting to
understand how to obtain realizations that make better use of the
ontology and to study setups where it can be unavoidable to exploit
the ontology in realizations. This is the case, for example, when
source queries are formulated in Datalog, the ontology is formulated
in (some extension of) \EL, and target queries are UCQs. Finally, we
note that it would be natural to study maximally contained
realizations instead of exact ones and to take into account
constraints over the source databases.

\subsection*{Acknowledgments}

We thank the anonymous reviewers for their helpful remarks. This
research was supported by ERC consolidator grant 647289 CODA.

\bibliographystyle{aaai}
\bibliography{references}

\clearpage

\ifextended

\appendix

\section{Appendix}

\section{Proof of Lemma~\ref{lem:patterns}}

\lempatterns*

\noindent
\begin{proof}
  We prove all statements for CQs, it is straightforward to generalize
  to UCQs: one only needs to employ the characterization of UCQ in
  terms of homomorphisms instead of the one for CQs.
\begin{enumerate}

\item Since $q_1 \subseteq q_2$, we have $q_2 \rightarrow q_1$. Let
  $h$ be a homomorphism witnessing this. It can be verified that
  restricting $h$ to the variables in the CQ
  $\Mbf(q_2)$\footnote{Including all the answer variables, which are
    the answer variables from $q_2$, even when they do not occur in
    any atom of $\Mbf(q_2)$.} yields a homomorphism from CQ
  $\Mbf(q_2)$ to CQ $\Mbf(q_1)$, thus $\Mbf(q_1)
  \subseteq_{\mn{sch}(\Mbf)} \Mbf(q_2)$. In fact, let $R(\xbf)$ be a
  relational atom in $\Mbf(q_2)$. Then $R(\xbf)$ was produced by some
  mapping $\varphi(\ybf) \rightarrow R(\zbf) \in \Mbf$ and
  homomorphism $h'$ from $\varphi(\ybf)$ to $q_2$ with
  $h'(\zbf)=\xbf$. Composing $h'$ with $h$ enables an application of
  the same mapping in $q_1$ that delivers $R(h(\xbf)) \in \Mbf(q_1)$,
  as required. Moreover, every equality atom $x=y$ in $\Mbf(q_2)$ must
  also be in $q_2$ by construction of $\Mbf(q_2)$.  Thus $h(x)=h(y)
  \in q_1$ or $h(x)=h(y)$.  In the former case, $h(x)=h(y) \in
  \Mbf(q_1)$ by construction of $\Mbf(q_1)$.

%\item Let $q_1 \subseteq_{\Sbf} q_2$, so there is a homomorphism $h:q_2 \rightarrow q_1$. We need to show that there exists a homomorphism $g:M(q_2) \rightarrow M(q_1)$. This homomorphism $g$ is given by $h$, restricted to the variables that occur in $M(q_2)$. Let $m(a_1,\ldots,a_n) \in M(q_2)$. Then there is a match from the body of $m$ to $q_2$. By $h$, there is also a match of the body of $m$ to $q_1$ and thus, $m(g(a_1), \ldots, g(a_n)) \in M(q_1)$.

\item From $r_1 \subseteq_{\mn{sch}(\Mbf)} r_2$, we obtain $r_2 \to
  r_1$. Let $h$ be a witnessing homomorphism. We need to show that for
  each CQ $p_1$ in the UCQ $\Mbf^-(r_1)$, there is a CQ $p_2$ in the
  UCQ $\Mbf^-(r_2)$ such that $p_2 \rightarrow p_1$. Thus let $p_1$ be
  from $\Mbf^-(r_1)$. By construction of $\Mbf^-(r_1)$, $p_1$ is
  obtained from $r_1$ by choosing a mapping from \Mbf for each
  relational atom in $r_1$, `applying it backwards', and then readding
  all equality atoms. We identify $p_2$ by choosing for every atom
  $R(\ybf)$ of $r_2$ the mapping chosen for $R(h(\ybf)) \in r_1$ in
  the construction of $p_1$. We can then straightforwardly define a
  homomorphism $h' $ from $p_2$ to $p_1$ by extending $h$ to the fresh
  variables in $p_2$.

%\item Let $r_1 \subseteq_{\Mbf} r_2$, so there is a homomorphism $h:r_2 \rightarrow r_1$. We need to show that there exists a homomorphism $g:M^-(r_2) \rightarrow M^-(r_1)$. Let $g$ be equal to $h$ on all variables that already occur in $r_2$. If $x$ is a fresh variable in $M^-(r_2)$, then $x$ corresponds to a non-answer-variable $y$ of the body of some mapping $m$ and some match of $m$ in $r_2$. Since $h:r_2 \rightarrow r_1$, $m$ also has a match in $r_1$ and in $M^-(r_1)$, a fresh variable $x'$ is introduced for $y$. We then set $g(x) = x'$.

\item For the ``only if'' direction, assume $p \rightarrow q$ for some
  CQ $p$ in the UCQ $\Mbf^-(r)$ and let $h$ be a witnessing
  homomorphism. Let $h'$ be the restriction of $h$ to the variables in
  $r$. It can be verified that $h'$ is a homomorphism from $r$ to
  $\Mbf(q)$. In fact, let $R(\xbf)$ be a relational atom in $r$. Then
  by construction of $\Mbf^-(r)$ there is a mapping $\varphi(\ybf)
  \rightarrow R(\zbf) \in \Mbf$ that was `applied backwards' in the
  construction of $p$ and thus there is a homomorphism $g$ from
  $\varphi(\ybf)$ to $\Mbf^-(r)$ with $g(\zbf)=\xbf$. Composing $g$
  with $h$ enables an application of the same mapping in $q$ that
  delivers $R(h(\xbf)) \in \Mbf(q)$, as required. The equality atoms
  in $r$ must also be satisfied since they are the same as in
  $\Mbf^-(r)$. % because every relational atom $R(\ybf)$ in $r$ is replaced in $p$
  % with the body $\exists \xbf \varphi(\ybf, \xbf)$ of some mapping,
  % which has a homomorphic image $\exists \xbf (h(\ybf), \xbf)$ in $q$,
  % and thus gets replaced with $R(\ybf)$ in $M(q)$.

  For the ``if'' direction, assume that there is a homomorphism $h$
  from $r$ to $\Mbf(q)$. We need to show that there is a CQ $p$ in the
  UCQ $\Mbf^-(r)$ such that $p \rightarrow q$. By construction, every
  atom $R(\xbf)$ in $\Mbf(q)$ is produced by some mapping
  $\varphi(\ybf) \rightarrow R(\zbf) \in \Mbf$ and homomorphism $g$
  from $\varphi(\ybf)$ to $q$ with $g(\zbf)=\xbf$. We identify
  $p$ by choosing for every atom $R(\ybf)$ of $r$ the mapping that
  produced $R(h(\ybf)) \in \Mbf(q)$. We can then straightforwardly
  define a homomorphism $h' $ from $p$ to $q$ by extending $h$ to
  the fresh variables in $p$.
\end{enumerate}
\end{proof}
\begin{lemma}
\label{lem:doublebackchain}
  Let $Q=(\Omc,\Sbf,q)$ be an OMQ with \Omc formulated in FO$(=)$,
  $q$ a UCQ, and $q_r$ an infinitary UCQ rewriting of $Q$. Then 
  $(\Omc,\Sbf,q_r) \subseteq_\Sbf q_r$.
\end{lemma}

\noindent
\begin{proof}
  For brevity, let $Q'=(\Omc,\Sbf,q_r)$.
% and $q_s'_r$ a rewriting of $Q'$. 
  Take an \Sbf-ABox \Amc and an $\abf \in \mn{adom}(\Amc)$ such that
  $\abf \in \mn{cert}_{Q'}(\Amc)$. 
% Since $q_s'_r$ is a rewriting of
%   $Q'$, this yields $\abf \in \mn{cert}_{Q'}(\Amc)$. Let $Q'' =
%   (\Omc,\Sbf,q_r)$. Since $p$ is the query in $Q'$ and a disjunct of
%   $q_r$, we obtain $\abf \in \mn{cert}_{Q''}(\Amc)$.
%
  We show that $\abf \in \mn{cert}_{Q}(\Amc)$, which implies $\abf
  \in \mn{ans}_{q_r}(\Amc)$ as desired since $q_r$ is a rewriting of
  $Q$.  Let \Imc be a model of $\Amc$ and \Omc. We have to show that
  $\abf \in \mn{ans}_{q}(\Imc)$.

  From $\abf \in \mn{cert}_{Q'}(\Amc)$, we obtain $\abf \in
  \mn{ans}_{q_r}(\Imc)$. This clearly implies that there is an
  interpretation $\Imc_f$ that is obtained by restricting \Imc to a finite
  subset of the domain and satisfies $\abf \in
  \mn{ans}_{q_r}(\Imc_f)$. Let $\Amc_\Imc$ be $\Imc_f$ viewed as an
  ABox, restricted to the symbols in \Sbf. Since $q_r$ uses only
  symbols from \Sbf, $\abf \in \mn{ans}_{q_r}(\Amc_\Imc)$. As $q_r$
  is a rewriting of $Q$, $\abf \in \mn{cert}_Q(\Amc_\Imc)$. Observe
  that \Imc is a model of \Omc and $\Amc_\Imc$, and thus it follows
  that $\abf \in \mn{ans}_{q}(\Imc)$, as required.
\end{proof}

% {\color{blue} On inconsistency: We then need to prove all our
% characterizations and lemmas for restricted classes of databases,
% writing e.g.\ ``$q_s \equiv_\Sbf \Mbf^-(q_r)$ relative to the class of
% all \Sbf-databases $D$ such that $\Mbf(D)$ is satisfiable w.r.t.\
% \Omc.'' Urgh.}

\section{Proof Details for DL-Lite}

\lemcanonicalrewriting*

\noindent
\begin{proof}\ 
 We show that $q_c \equiv_{\mn{sch}(\Mbf)} q_r$. Since $q_r$ is a
UCQ-rewriting of $Q$, it then follows that $q_c$ is also a
UCQ-rewriting of $\Omc$. We consider the two directions of
the equivalence separately.

$q_c \supseteq_{\mn{sch}(\Mbf)} q_r$. This holds because every CQ $q$ in
$q_r$ is also an element in $q_c$. In fact, $q(\xbf)$ being in $q_r$
means that it is of size at most $n$. Viewing $q$ as an ABox and \xbf
as a candidate answer, we trivially have $\xbf \in \mn{ans}_{q_r}(q)$
and thus $\xbf \in \mn{cert}_Q(q)$ because $q_r$ is a rewriting of
$Q$. As a consequence, the pair $(q,\xbf)$ gives rise to a CQ
in $q_c$.

$q_c \subseteq_{\mn{sch}(\Mbf)} q_r$. Let $q$ in $q_c$. We have to
show that there is a CQ $q'$ in $q_r$ such that $q' \to q$. By
definition of $q_c$, $q$ is a pair $(\Amc,\abf)$ viewed as a CQ such
that $\abf \in \mn{cert}_Q(\Amc)$. Because $q_r$ is a rewriting of
$Q$, it follows from the latter that
$\abf \in \mn{ans}_{q_r}(\Amc)$. This means that there is some $q'$ in
$q_r$ with a homomorphism $h$ from $q'$ to \Amc that maps the answer
variables of $q'$ to $\abf$. As $q$ is just $\Amc$ with $\abf$ as the
answer variables, $h$ shows $q' \to q$.
\end{proof}
\thmpiptwolower*

In preparation for the proof of Theorem~\ref{thm:pip2lower}, let us
recall the standard representation of 3SAT as a constraint
satisfaction problem (CSP) and sketch how this can be used to show
that the CQ-to-CQ expressibility problem is $\NPclass$-hard for GAV
mappings and the empty ontology. Let $\varphi(y_1,\ldots,y_m)$
% \[
% \varphi = \exists y_0\cdots\exists y_m
% \psi(x_1,\ldots,x_n,y_1,\ldots,y_m)
% \]
be a propositional logic formula in 3CNF. Let \Sbf be the schema that
consists of all ternary relation names $C_{u_1u_2u_3}$ with $u_1u_2u_3 \in
\{ \mathsf{n}, \mathsf{p} \}^3$. Every clause in $\vp$ can be viewed
as a fact over signature $S$ by letting the $u_i$ represent the
polarities of the variables in the clause and using the variables from
$\varphi$ as constants. For example, $\neg y_2 \vee y_1 \vee \neg y_3$
gives the fact
$C_{\mathsf{n},\mathsf{p},\mathsf{n}}(y_2,y_1,y_3)$. Thus, $\vp$ can
be viewed as a database $D_\vp$. What's more, we can build a database
$D$ that is independent of $\vp$ and such that $D_\vp \rightarrow D$
if and only if $\vp$ is satisfiable. In CSP, $D$ is called the
\emph{template for 3SAT}. It is actually easy to find $D$: use two
constants 0 and 1 that represent truth values and add the fact
$C_{u_1u_2u_3}(t_1,t_2,t_3)$ if the truth assignment $t_1,t_2,t_3$ to
the three variables of a clause with polarities $u_1u_2u_3$ makes the
clause true. For example,
$C_{\mathsf{n},\mathsf{p},\mathsf{n}}(t_1,t_2,t_3)$ is added for any
$t_1t_2t_3 \in \{0,1\}^3$ except~$101$.

How does this relate to the expressibility problem? Let $q_s$ be
$D_\vp$ viewed as a Boolean CQ and take the mappings $D_\vp
\rightarrow A(x)$ and $D \rightarrow A(x)$ where $D_\vp$ and $D$ are
viewed as CQs in which all variables are answer variables with $x$ an
arbitrary but fixed such variable. Then $\Mbf^-(\Mbf(q_s))$ is
(equivalent to) $D_\vp \vee D$ with $D_\vp$ and $D$ viewed as a
Boolean CQs and thus it remains to apply Theorem~\ref{thm:maincharact}
where $q_r$ is now simply $\Mbf(q_s)$ since the ontology is empty.

\medskip

We now lift this simple reduction to $\forall \exists$-3SAT. Thus let
\[
 \varphi = \forall x_0\cdots \forall x_n \exists y_0\cdots\exists y_m
\psi(x_1,\ldots,x_n,y_1,\ldots,y_m)
\]
be a quantified Boolean formula with $\psi$ in 3CNF. We construct an
OBDA specification $(\emptyset, \Mbf, \Sbf)$ with \Mbf a set of GAV
mappings as well as a Boolean CQ $q_s$ over schema $\Sbf$ such that
$\varphi$ is true iff $q_s$ is CQ-expressible in $(\emptyset, \Mbf,
\Sbf)$.  

The universally quantified variables have a different status in the
reduction as, unlike the existentially quantified variables, they are
not represented by variables in~$q_s$. For example, the clause $(y_1
\vee \neg x_0 \vee \neg y_1)$ gives rise to the atom $C_{\mn{p} \neg
  x_0 \mn{n}}(y_1,y_1)$. This is compensated by constructing the
mappings in \Mbf so that $\Mbf^-(\Mbf(q_s))$ is now essentially a
disjunction of ($q_s$ and) exponentially many versions of the template
$D_\varphi$, one for every truth assignment to the universally
quantified variables. For example, such a template includes $C_{\mn{p}
  \neg x_0 \mn{n}}(t_1t_2)$ for \emph{all} $t_1t_2 \in \{0,1\}^2$ if
it represents a truth assignment that makes $x_0$ true and otherwise
it includes $C_{\mn{p} \neg x_0 \mn{n}}(t_1t_2)$ for all $t_1t_2 \in
\{0,1\}^2$ except~01. To achieve this, we use binary relations in the
head of mappings instead of unary ones. In fact, we want $\Mbf(q_s)$
to be of the form $\Mbf(q_s) = \bigwedge_{i=0}^n r_i(y_0,y_1)$ and
there will be two ways to translate each $r_i(y_0,y_1)$ backwards in
the construction of $\Mbf^-(\Mbf(q_s))$, corresponding to the two
possible truth values of the universally quantified variable $x_i$.

We now make the reduction precise.  Let $U = \{x_0,\ldots,x_n,\neg
x_0, \ldots, \neg x_n, \mn{n}, \mn{p}\}$. For every triple
$(u_1,u_2,u_3) \in U^3$ we include a relation $C_{u_1u_2u_3}$
in~$\Sbf$. The \emph{arity} of $C_{u_1u_2u_3}$ is the number of
positions in $(u_1, u_2, u_3)$ that are $\mn{n}$ or
$\mn{p}$. Additionally, $\Sbf$ contains a binary relation $Z$ which
helps us to achieve that $\Mbf(q_s)$ is of the intended form even when
$q_s$ admits non-trivial automorphisms.

We define $q_s$ to encode $\varphi$. The existentially quantified
variables of $q_s$ are $y_0,\ldots,y_m$. For every clause $\ell_1 \vee
\ell_2 \vee \ell_3$ in $\psi$, we introduce an atom in $q_s$ with the
symbol $C_{u_1 u_2 u_3}$, where $u_i=\ell_i$ if $\ell_i$ contains a
universally quantified variable, $u_i=\mn{p}$ if $\ell_i$ is a
positive literal with an existentially quantified variable and
$u_i=\mn{n}$ if $\ell_i$ is a negative literal with an existentially
quantified variable. The variables that occur in this atom are the
existentially qualified variables of the clause in the order of their
appearance in the clause, see above for an example.  Moreover, we add
the atom $Z(y_0,y_1)$ to $q_s$, assuming w.l.o.g.\ that there are at
least two existentially quantified variables in $\varphi$.

We now construct the GAV mappings in \Mbf. 
For every universally quantified variable $x_i$ of $\varphi$, we introduce
three mappings with the same head $r_i (z_0,z_1)$:
\begin{enumerate}
 \item In the first mapping $q_s'(z_0,z_1)
\rightarrow r_i(z_0,z_1)$, the body is $q_s$ with $y_0$ and $y_1$ 
renamed to $z_0,z_1$ (and all existential quantifiers removed).
\item \label{x is false} The body of the second mapping
  $\tau^0_i(z_0,z_1) \rightarrow r_i(z_0,z_1)$ generates the
  part of the template that must be there when $x_i$ is assigned
  truth value $0$. The variables $z_0$ and $z_1$ represent the two
  elements of the template.

  The body $\tau^0_i$ only contains the variables $z_0$ and $z_1$. For
  every $u_1u_2u_3 \in U^3$ and sequence $\overline{v}=v_1v_2\cdots$
  over $\{z_0,z_1\}$ of the same length as the arity of $C_{u_1 u_2
    u_3}$, we add the atom $C_{u_1 u_2 u_3}(\overline{v})$ to
  $\tau^0_i$ if at least one of the following holds:
\begin{enumerate}
 \item $\neg x_i$ is among $u_1$, $u_2$ and $u_3$, \label{x makes true}
\item $v_i = z_0$ and the $i$-th appearance of $\mn{n}$ or $\mn{p}$ in
$(u_1,u_2,u_3)$ is $\mn{n}$ for some $i$,
 \item $v_i = z_1$ and the $i$-th appearance of $\mn{n}$ or $\mn{p}$ in
$(u_1,u_2,u_3)$ is $\mn{p}$ for some $i$.
\end{enumerate}
We also add the atoms $Z(z_0,z_0)$, $Z(z_0,z_1)$, $Z(z_1,z_0)$ and  
$Z(z_1,z_1)$ to $\tau^0_i$.  
 
(For example in $\tau^0_3$ we add the atoms $C_{\mn{p} x_3
  \mn{n}}(z_0,z_0)$, $C_{\mn{p} x_3 \mn{n}}(z_1,z_0)$, and $C_{\mn{p}
  x_3 \mn{n}}(z_1,z_1)$, but not $C_{\mn{p} x_3 \mn{n}}(z_0,z_1)$.  We
add all $C_{\mn{p} \neg x_3 \mn{n}}(z_0,z_0)$, $C_{\mn{p} \neg x_3
  \mn{n}}(z_0,z_1)$, $C_{\mn{p} \neg x_3 \mn{n}}(z_1,z_0)$ and
$C_{\mn{p} \neg x_3 \mn{n}}(z_1,z_1)$.  Also, we add $C_{x_2 x_3
  \mn{p}} (z_1)$ but not $C_{x_2 x_3 \mn{p}} (z_0)$.)
 \item \label{x is true} The body $\tau^1_i(z_0,z_1)$ of the third mapping
$\tau^1_i(z_0,z_1) \rightarrow r_i(z_0,z_1)$ is dual to
$\tau^0_i(z_0,z_1)$ in that it encodes the case where $x_i$ is true. This
means the definition is as above with the difference that in
Condition~\ref{x makes true}, the literal $x_i$, and not $\neg x_i$, is
required to be among $u_1$, $u_2$ and $u_3$.
\end{enumerate}
\begin{lemma}
$\varphi$ is true iff $q_s$ is CQ-expressible in $(\emptyset, \Mbf, \Sbf)$. 
\end{lemma}
\noindent  
\begin{proof}
  By Theorem~\ref{thm:maincharact}, it suffices to show that $\varphi$
  is true iff $\Mbf^-(\Mbf(q_s)) \subseteq q_s$, that is, iff there is
  a homomorphism from $q_s$ to every disjunct of $\Mbf^-(\Mbf(q_s))$.

  We first describe the UCQ $\Mbf^-(\Mbf(q_s))$. First observe that
  $\Mbf(q_s)$ is indeed $\bigwedge_{i=0}^n r_i(y_0,y_1)$: The mapping
  $q_s' \rightarrow r_i(z_0,z_1)$ has a match at $(y_0,y_1)$ for every
  $i$ and it has no other matches since $Z(z_0,z_1)$ is in $q_s'$ and
  $Z(y_0,y_1)$ is the only atom in $q_s$ the contains $Z$. The
  mappings $\tau_i^v(z_0, z_1) \rightarrow r_i(z_0, z_1)$ do not match
  anywhere in $q_s$ for $v \in \{0, 1\}$, since the atoms $Z(z_0,
  z_0)$, $Z(z_0, z_1)$, $Z(z_1, z_0)$ and $Z(z_1, z_1)$ all appear in
  the body of these mappings. There are $3^{n+1}$ disjuncts in
  $\Mbf^-(\Mbf(q_s))$, one for every combination of $n+1$ choices of
  the three different mappings with head $r_i(z_0, z_1)$, for every $i
  \in \{0,\ldots,n\}$.
We now prove the lemma.

\smallskip

``$\Rightarrow$''. Assume that $\varphi$ is true. We want to show
that there is a homomorphism from $q_s$ into every CQ in
$\Mbf^-(\Mbf(q_s))$. Pick an arbitrary CQ $p$ in
$\Mbf^-(\Mbf(q_s))$. Such a disjunct corresponds of a choice of one of
the bodies $q_s'$, $\tau^0_i$, or $\tau^1_i$ for each $i \in \{0,\dots,n\}$.

First consider the case where for some $i$ we choose $q_s'$. In that
case $p$ contains an isomorphic copy of $q_s$ and thus we are done.

Now consider the case for no $i$ we choose a mapping with body $q_s'$,
that is, for every $i$ we choose $\tau^0_i$ or~$\tau^1_i$. This
corresponds to an assignment $t$ of the truth values $0$ and $1$ to
the variables $x_0,\dots,x_n$. Because $\varphi = \forall
x_0,\dots,x_n \exists y_0, \dots, y_m \psi$ is true we can extend $t$
to an assignment for $x_0,\dots,x_n,y_0,\dots,y_m$ that makes $\psi$
true. We define the homomorphism $h$ from $q_s$ to $p$ such that
$h(y_j) = y_{t(y_j)}$ for all $j \in \{0,\dots,m\}$. We need to
verify that $h$ is a homomorphism. All atoms with the symbol $Z$ are
preserved as by definition of the $\tau^0_i$ and $\tau^1_i$, there is
a $Z$ atom in $p$ for any pair over $\{y_0,y_1\}$. Consider then any
atom $C_{u_1 u_2 u_3} (\overline{y})$ from $q_s$.  There is a
corresponding clause $\ell_1 \lor \ell_2 \lor \ell_3$ in $\psi$ that
is true under the assignment $t$. It follows that one of the literals
$\ell_1,\ell_2,\ell_3$ is true under $t$. We make a case distinction:
% distinguish cases
% depending on whether $\ell_k$ is a positive or negative and on whether
% it is a literal for a universally or an existentially quantified
% variable.

If $\ell_k = x_i$ is a universally quantified variable, then $t(x_i) =
1$ and hence $p$ contains $\tau^1_i$. By (the implicit) Condition~3a,
from the definition of \Mbf, it follows that $\tau^1_i$ contains the
atom $C_{u_1 u_2 u_3} (y_{t(\overline{y})})$.

In the case where $\ell_k = \neg x_i$, we use the same argument and
Condition~\ref{x makes true}.

If $\ell_k = y_j$ is an existentially quantified variable, then $u_k =
\mn{p}$ and $t(y_j) = 1$. Hence, $h(y_j) = y_1$ and from
Condition~(2c) or~(3c), the atom $C_{u_1u_2u_3}$ is in $\tau^0_0$ and
in $\tau^1_0$, thus in $p$ (since we can assume w.l.o.g. that there
is at least one universally quantified variable).

The case where $\ell_k = \neg y_j$ is analogous, using Conditions~(2b)
or (3b).

\smallskip

``$\Leftarrow$''. Assume that there is a homomorphism from $q_s$ into
every disjunct of $\Mbf^{-}(\Mbf(q_s))$. We want to show that
$\varphi$ is true. So take any assignment $t$ for $x_0,\dots,x_n$. We
need to extend $t$ to an assignment $t'$ for
$x_0,\dots,x_n,y_0,\dots,y_m$ that makes $\psi$ true. Consider the
disjunct $p_t$ of $\Mbf^{-}(\Mbf(q_s))$ that arises from choosing the
mapping $\tau^{t(x_i)}_i(z_0,z_1) \to r_i(z_0,z_1)$ for each $i =
0,\dots,n$. By assumption, there is a homomorphism $h$ from $q_s$ to
$p_t$. Clearly, $p_t$ contains only the 
variables $y_0$ and $y_1$. We define $t'$ such that $t'(x_i) =
t(x_i)$, $t'(y_j) = 0$ if $h(y_j) = y_0$, and $t'(y_j) = 1$ if $h(y_j)
= y_1$.

It remains to be shown that $\psi$ is true under $t'$. Take an
arbitrary clause $\ell_1 \lor \ell_2 \lor \ell_3$ in $\psi$. Consider
the corresponding atom $C_{u_1 u_2 u_3}(\overline{y})$ in $q_s$. As
$h$ is a homomorphism, $C_{u_1 u_2 u_3}(h(\overline{y}))$ is in
$p_t$. By the definition of $p_t$ this means that $C_{u_1 u_2
  u_3}(h(\overline{y}))$ is contained in $\tau^{t(x_i)}_i$ for
some~$i$. Hence one of the conditions (a), (b) or (c) from Point~2
or~3 of the construction of \Mbf is satisfied.

If (a) is satisfied and $t(x_i) = 0$, then $\neg x_i$ is among
$u_1,u_2,u_3$. It follows that then $\neg x_i$ is a literal in $\ell_1
\lor \ell_2 \lor \ell_3$ and hence the clause is true under $t'$ because
$t'(x_i) = t(x_i)$. In case (a) is satisfied and $t(x_i) = 1$, we
can reason analogously.

If (b) is satisfied, then $h(y_i) = y_0$ for some $y_i$ in
$\overline{y}$, and the $i$-th appearance of either $\mn{n}$ or $\mn{p}$ in
$(u_1,u_2,u_3)$ is $\mn{n}$. Hence $\neg y_i$ is a literal in $\ell_1 \lor \ell_2
\lor \ell_3$ which makes the clause true because $t'(y_1) = 0$ since
$h(y_i) = y_0$. We reason analogously in the case where (c) is
satisfied.
\end{proof}

\section{Preliminary: Two-way alternating parity automata (TWAPA)}

We introduce two-way alternating parity automata on finite trees
(TWAPAs). 

A \emph{tree} is a non-empty (and potentially infinite) set $T \subseteq
\mathbb{N}^*$ closed under prefixes. We say that $T$ is \emph{$m$-ary}
if $T \subseteq \{1,\dots,m\}^*$. For an alphabet $\Gamma$, a
\emph{$\Gamma$-labeled tree} is a pair $(T,L)$ with $T$ a tree and $L:T
\rightarrow \Gamma$ a node labeling function.

%
 % Let \Nbbm\xspace denote the \emph{positive} integers. A
% \emph{tree} is a non-empty (and potentially infinite) set $T \subseteq
% \Nbbm^*$ closed under prefixes.
%  such that if $x \cdot i \in T$ with $x \in
% \Nbbm^+$ and $c \in \Nbbm$, then $x \cdot i' \in T$ whenever $0 < i' <
% i$.
% The node $\varepsilon$ is the \emph{root} of $T$. As a convention, we
% take $x \cdot 0 = x$ and $ (x \cdot c) \cdot -1 = x$. Note that
% $\varepsilon \cdot -1$ is undefined.  We say that $T$ is
% \emph{$m$-ary} if for every $x \in T$, the set $\{ i \mid x \cdot i
% \in T \}$ is of cardinality at most $m$.  W.l.o.g., we assume that all
% nodes in an $m$-ary tree are from $\{1,\dots,m\}^*$.
%We use $[m]$ to denote the set $\{-1,0,\dots,m\}$ and f
For any set $X$,
let $\Bmc^+(X)$ denote the set of all positive Boolean formulas over
$X$, i.e., formulas built using conjunction and disjunction over the
elements of $X$ used as propositional variables, and where the special
formulas $\mn{true}$ and $\mn{false}$ are allowed as well. 
% For an
% alphabet $\Gamma$, a \emph{$\Gamma$-labeled tree} is a pair $(T,L)$
% with $T$ a tree and $L:T \rightarrow \Gamma$ a node labeling function.
An \emph{infinite path} $P$ of a tree $T$ is a
prefix-closed set $P \subseteq T$ such that for every $i \geq 0$,
there is a unique $x \in P$ with $|x|=i$.

\begin{definition}[TWAPA]
  A \emph{two-way alternating parity automaton 
    (TWAPA) on finite $m$-ary trees} is a tuple
  $\Amf=(S,\Gamma,\delta,s_0,c)$ where $S$ is a finite set of
  \emph{states}, $\Gamma$ is a finite alphabet, $\delta: S \times
  \Gamma \rightarrow \Bmc^+(\mn{tran}(\Amf))$ is the \emph{transition
    function} with $\mn{tran}(\Amf) = \{ \langle i \rangle s, \ [i] s
  \mid -1 \leq
  i \leq m \text{ and } s \in S \}$ the set of
  \emph{transitions} of \Amf, $s_0 \in S$ is the \emph{initial state},
  and $c:S \rightarrow \mathbb{N}$ is the \emph{parity condition} that 
  assigns to each state a \emph{priority}.
\end{definition}
Intuitively, a transition $\langle i \rangle s$ with $i>0$ means that
a copy of the automaton in state $s$ is sent to the $i$-th successor
of the current node, which is then required to exist. Similarly,
$\langle 0 \rangle s$ means that the automaton stays at the current
node and switches to state $s$, and $\langle -1 \rangle s$ indicates
moving to the predecessor of the current node, which is then required
to exist. Transitions $[i] s$ mean that a copy of the automaton in
state $s$ is sent on the relevant successor if that successor exists
(which is not required).
\begin{definition}[Run, Acceptance]
  A \emph{run} of a TWAPA $\Amf = (S,\Gamma,\delta,s_0,c)$ on a finite
  $\Gamma$-labeled tree $(T,L)$ is a $T \times S$-labeled tree
  $(T_r,r)$ such that the following conditions are satisfied:
  \begin{enumerate}

  \item $r(\varepsilon) = ( \varepsilon, s_0)$;
    
  % \item if $y \in T_r$, $r(y)=(x,s)$, and $\delta(s,L(x))=\vp$, then
  %   there is a (possibly empty) set $S = \{ (c_1,s_1),\dots,(c_n,s_n)
  %   \} \subseteq \mn{tran}(\Amf)$ such that $S$ satisfies $\vp$ and
  %   for $1 \leq i \leq n$, $x \cdot c_i$ is defined and  a node
  %   in $T$, and there is a $y \cdot i \in T_r$ such that $r(y \cdot
  %   i)=(x \cdot c_i,s_i)$.

  \item if $y \in T_r$, $r(y)=(x,s)$, and $\delta(s,L(x))=\vp$, then
    there is a (possibly empty) set $S \subseteq \mn{tran}(\Amf)$ such
    that $S$ (viewed as a propositional valuation) satisfies $\vp$ as
    well as the following conditions:
    \begin{enumerate}

    \item if $\langle i \rangle s' \in S$, then $x \cdot i \in T$ and 
      there is a node $y \cdot j \in T_r$ such that $r(y \cdot j)=(x 
      \cdot i,s')$;

    \item if $[i]s' \in S$ and $x \cdot i \in T$, then
      there is a
      node $y \cdot j \in T_r$ such that $r(y \cdot j)=(x \cdot
      i,s')$.

    \end{enumerate}

  \end{enumerate}
  We say that $(T_r,r)$ is \emph{accepting} if on all infinite paths
  $\varepsilon = y_1 y_2 \cdots$ of $T_r$, the maximum priority that
  appears infinitely often is even.  A finite $\Gamma$-labeled tree
  $(T,L)$ is \emph{accepted} by \Amf if there is an accepting run of
  \Amf on $(T,L)$. We use $L(\Amf)$ to denote the set of all finite
  $\Gamma$-labeled tree accepted by \Amf.
\end{definition}
It is known (and easy to see) that TWAPAs are closed under
complementation and intersection, and that these constructions
involve only a polynomial blowup.
%
% , i.e., for any TWAPA \Amf over finite
% $\Gamma$-labeled $m$-ary trees, there is a TWAPA $\overline{\Amf}$
% over finite $\Gamma$-labeled $m$-ary trees such that
% $L(\overline{\Amf})$ is the set of those finite $\Gamma$-labeled
% $m$-ary trees $(T,L)$ such that $(T,L) \notin L(\Amf)$.
It is also known
that % TWAPAs are closed under union and intersection, and that
their emptiness problem can be solved in time single exponential in
the number of states and polynomial in all other components of the
automaton. % Since our TWAPA will only use priorities 1 and 2, we can
% assume w.l.o.g.\ that the runtime of the emptiness test is independent
% of the acceptance condition. What's more,
In what follows, we shall generally only explicitly
analyze the number of states of a TWAPA, but only implicitly take care
that all other components are of the allowed size for the complexity
result that we aim to obtain.

\section{Preliminary: Universal models}

We introduce the universal model $\Imc_{\Amc,\Omc}$ of an ABox $\Amc$ and an ontology $\Omc$ in
$\ELHI$. The main properties of $\Imc_{\Amc,\Omc}$ are:
\begin{itemize}
\item $\Imc_{\Amc,\Omc}$ is a model of $\Amc$ and $\Omc$;
\item for every model $\Imc$ of $\Amc$ and $\Omc$ there exists a homomorphism
from $\Imc_{\Amc,\Omc}$ to $\Imc$ that maps each $a \in \mn{adom}(\Amc)$ to
itself.
\end{itemize}
$\Imc_{\Amc,\Omc}$ is constructed using a standard chase procedure. 
We assume that $\Omc$ is in normal form.

We start by defining the universal model $\Imc_{\Amc,\Omc}$ of $\Amc$ and $\Omc$.
It is convenient to use ABox notation when constructing $\Imc_{\Amc,\Omc}$ and so we will
construct a (possibly infinite) ABox $\Amc^{\text{uni}}_{\Omc}$ and define $\Imc_{\Amc,\Omc}$ as the
interpretation corresponding to $\Amc_{\Omc}^{\text{uni}}$.

Thus assume that $\Amc$ and $\Omc$ are given. The \emph{full completion sequence of $\Amc$ w.r.t.~$\Omc$}
is the sequence of ABoxes $\Amc_0,\Amc_1,\dots$ defined by setting $\Amc_0 = \Amc$ and defining $\Amc_{i+1}$ to be $\Amc_i$ extended as follows (recall that we abbreviate $r(a,b)$ by $r^{-}(b,a)$
and that $r$ ranges over roles):
\begin{itemize}

\item[(i)] If $\exists r.B \sqsubseteq A \in \Omc$ and $r(a,b), B(b) \in \Amc_i$, then add $A(a)$ to $\Amc_{i+1}$;

%\item[(ii)] for each $\exists r^- . B \sqsubseteq A \in \Omc$ such that 
%$r(b,a), B(b) \in \Amc_i$ and $A(a) \notin \Amc_i$, add $A(a)$ to $\Amc_{i}$;

\item[(ii)] if $\top \sqsubseteq A \in \Omc$ and $a \in \mn{adom}(\Amc_{i})$,
then add $A(a)$ to $\Amc_{i+1}$;

\item[(iii)] if $B_{1}\sqcap B_{2} \sqsubseteq A\in \Omc$ and $B_{1}(a),B_{2}(a)\in \Amc_{i}$, 
then add $A(a)$ to $\Amc_{i+1}$;
%\item[(iii)] for each $A(a)$ such that $\Amc_i|_a,\Omc\models A(a)$ and
%  $A(a) \notin \Amc_i$, add $A(a)$ to $\Amc_{i}$;

\item[(iv)] if $A \sqsubseteq \exists r.B \in \Tmc$ and $A(a) \in \Amc_i$ then take a fresh individual $b$ and add $r(a,b)$ and $B(b)$ to $\Amc_{i+1}$;

\item[(v)] if $r\sqsubseteq s \in \Omc$ and $r(a,b) \in \Amc_{i}$, then add $s(a,b)$ to $\Amc_{i+1}$.
\end{itemize}
Now let $\Amc_{\Omc}^{\text{uni}}=\bigcup_{i\geq 0}\Amc_{i}$ and let $\Imc_{\Amc,\Omc}$ be the interpretation
corresponding to $\Amc_{\Omc}^{\text{uni}}$. It is straightforward to prove the following properties
of $\Imc_{\Amc,\Omc}$.
\begin{lemma} \label{lem:canmodelproperties}
Assume $\Omc$ is in normal form.
Then
\begin{itemize}
\item $\Imc_{\Amc,\Omc}$ is a model of $\Amc$ and $\Omc$;
\item for every model $\Imc$ of $\Amc$ and $\Omc$ there exists a homomorphism
from $\Imc_{\Amc,\Omc}$ to $\Imc$ that maps each $a\in \mn{adom}(\Amc)$ to
itself.
\end{itemize}
\end{lemma}
Note that the ABox $\Amc_{\Omc}^{\text{uni}}$ can contain additional
constants and can even be infinite.

The proof of the following is straightforward.
\begin{lemma}
\label{lem:complabox}
For all facts $A(a)$ and $r(a,b)$ with $a,b\in \mn{adom}(\Amc)$:
\begin{itemize}
\item $\Amc,\Tmc \models A(a)$ iff $A(a) \in \Amc^{\text{uni}}_{\Omc}$;
\item $\Amc,\Tmc \models r(a,b)$ iff $r(a,b) \in \Amc^{\text{uni}}_{\Omc}$. 
\end{itemize}
\end{lemma}

\section{Preliminary: Derivation Trees}

In the next sections we use TWAPA to obtain $\ExpTime$-decision
procedures for entailment of atomic queries. The construction of these
automata relies on a characterization of entailment of AQs in term of
derivation trees.

Fix an $\ELHI$ ontology \Omc in normal form and an ABox \Amc. A
\emph{derivation tree} for a fact $A_0(a_0)$ in \Amc, with $A_0 \in
\NC$ and $a_0 \in \mn{adom}(\Amc)$, is a finite $\mn{adom}(\Amc) \times
\NC$-labeled tree $(T,V)$ that satisfies the following conditions:
\begin{enumerate}

\item $V(\varepsilon)=(a_0,A_0)$;

% \item if $x \neq \varepsilon$, then $V(x)$ is not of the form
%   $(a,\bot)$;

% \item if $V(x)=(a,A)$ and $A(a) \in \Amc$ or $\top \sqsubseteq A \in
%   \Omc$, then $x$ is a leaf;

% \item if $V(x)=(a,A)$, then $x$ has one successor $y$ with $V(y)=(a,\Gamma)$
%   for some $\Gamma$ with $\Omc \models \Gamma \sqsubseteq A$;

\item if $V(x)=(a,A)$ and neither $A(a) \notin \Amc$ nor $\top
  \sqsubseteq A \in \Omc$, then one of the following holds:
  \begin{itemize}

  \item $x$ has successors $y_1,\dots,y_k$, $k \geq 1$ with
    $V(y_i)=(a,B_i)$ for $1 \leq i \leq k$ and $\Omc \models B_1
    \sqcap \cdots \sqcap B_k \sqsubseteq A$;
    % (then $x$ is of  \emph{local type});

  \item $x$ has a single successor $y$ with $V(y)=(b,B)$ and there
    is an $\exists s . B \sqsubseteq A \in \Omc$ and an $r(a,b) \in
    \Amc$ such that
    $\Omc \models r \sqsubseteq s$.
    % (then $x$ is of \emph{existential type}).

  \end{itemize}

\end{enumerate}
%
% Note that there might be more than one derivation tree for the same
% assertion $A_0(a_0)$. Also note that derivations trees are essentially
% the same thing as proof trees for datalog.
Note that the first item of Point~2 above requires $\Omc \models
A_1\sqcap \dotsb\sqcap A_n \sqsubseteq A$ instead of $A_1 \sqcap A_2
\sqsubseteq A \in \Omc$ to `shortcut' anonymous parts of the universal
model. In fact, the derivation of $A$ from $A_1 \sqcap \dots \sqcap
A_n$ by $\Omc$ can involve the introduction of anonymous elements.
\begin{lemma}
\label{lem:derivationtrees}
Let $A \in \NC$ and let $a$ be a constant in $\Amc$. We have $\Amc,\Omc \models A(a)$ iff there is
    a derivation tree for $A(a)$ in \Amc.

\end{lemma}
The proof is a straightforward variation of an analogous result for
the stronger logic $\mathcal{ELIHF}^{\cap-\mathsf{lhs}}_\bot$ in
\cite{BHLW-IJCAI16}. Details are omitted.

\section{Expressibility in \ELHI: Upper Bound for Rooted Queries,
  Missing Details}

% TODO: This restate does not really work properly.
\lemnexpalgcorr*

\noindent
\begin{proof}
As explained in the main text, the soundness part is easy.

For the completeness direction assume that $\Mbf^-(q_r) \subseteq_\Sbf
q_s$ is false. This means there is an ABox $\Amc$ and tuple $\abf$ such that
$(\Amc,\abf)$ is in the canonical rewriting $q_r$ of $Q$ and there is
$p$ in $\Mbf^-(\Amc,\abf)$ such that for all $q$ in $q_s$ we have $q
\not \to p$. We show how to obtain an ABox $\Amc''$ that the algorithm
can choose in step~\ref{guess abox} in order to accept.

By Lemma~\ref{lem:unravel} there exists an pseudo tree-shaped ABox
$\Amc'$ of core-size $|q_t|$ and branching degree at most $|\Omc|$ and
tuple $\abf'$ in the core of $\Amc'$ such that $\abf' \in
\mn{cert}_Q(\Amc')$ and there is a homomorphism $(\Amc',\abf') \to
(\Amc,\abf)$.

% is some $p'(\abf') \in \Mbf^-(\Amc')$ such that $q_s(\xbf) \not\to p'(\abf)$.

Define $\Amc''$ to be obtained from the
pseudo tree-shaped ABox $\Amc'$ by removing all facts that
contain at least one constant that has a distance larger than
$|q_s|$ from the core. Furthermore, we add all facts $A(b)$ and
$r(b,b)$ for all $A$ and $r$ in $\mn{sch}(\Mbf)$ and for all
constants $b$ that have exactly distance $|q_s|$ from the core.

The resulting size of $\Amc''$ is at most $|q_t| + |q_t| \cdot |\Omc|^{|q_s|}$, so $\Amc''$ can be chosen in Step~1.
\\[2mm]
 {\bf Claim:} Step~\ref{verify truth} succeeds, that is, $\abf' \in
\mn{cert}_Q(\Amc'')$.
\\[2mm]
\emph{Proof of claim:} 
Define a homomorphism $h : \Amc' \to \Amc''$ that is the identity on
constants of distance at most $|q_s|$ from the core and maps a constant
$b$ of distance more than $|q_s|$ from the core to the unique $b'$ of
distance precisely $|q_s|$ such
that $b$ is in the subtree rooted at $b'$. This indeed defines a
homomorphism because in $\Amc''$ all symbols in $\mn{sch}(\Mbf)$ are
true at each constant with distance $|q_s|$. Now $\abf' \in
\mn{cert}_Q(\Amc'')$ follows from $\abf' \in \mn{cert}_Q(\Amc')$
because $h$ is a homomorphism that fixes the tuple $\abf'$.
\\[2mm]
We then describe the query $p'$ in $\Mbf^-(\Amc'')$ that can be chosen in
step~\ref{choose disjunct}. For every fact in $\Amc''$ that is also in
$\Amc'$ choose the same mapping that was chosen to obtain $p$ from
$\Amc$. For all other facts we choose an arbitrary suitable mapping.
\\[2mm]
 {\bf Claim:} Step~\ref{verify no hom} succeeds, that is, $q
\not\to p'$ for all $q$ in $q_s$.
\\[2mm]
\emph{Proof of claim:} Assume towards contradiction that there is a
homomorphism $g' : q \to p'$ for some $q$ in $q_s$. Because $\Amc''$ and
$\Amc'$ are equal when restricted to variables of distance less than
$n$ from $\abf'$ it follows that $p$ and $p'$ are equal when
restricted to variables of distance less than $n$ from $\abf'$. This
is due to the fact that the distance of two variables from $\Amc$
cannot decrease in $p$ in $\Mbf^-(\Amc)$, and similarly for
$\Amc''$ and $p'$.

Because $|q| \leq |q_s|$ and $q$ is rooted it follows that all
constants in the image of $g' : q \to p'$ have distance less than $|q_s|$
from $\abf$ in $p$. Since $\abf$ lies in the core, these constants have also distance less than $|q_s|$ from the core. Because $p$ and $p'$ are equal when restricted to
constants of distance less than $|q_s|$ from the core, there is a
homomorphism $g : q \to p$, which contradicts our assumption on $p$.
\\[2mm]
\end{proof}

\section{Verification in \ELHI: Upper Bound, Missing Details}

We describe the construction of the TWAPA $\Amf^\lambda$. Recall that
$\Amf^\lambda$ will get as input (an encoding of) a triple $(\Amc,
\abf, p)$ consisting of a pseudo tree ABox $\Amc$ with a core $\Cmc$
of size at most $|q_t|$, a tuple $\abf$ of the same arity as $q_t$ and
a CQ $p \in \Mbf^-(\Amc)$, while $\lambda$ is a homomorphism pattern
that maps every variable $y$ in $q_t$ to a tuple $(c,o) \in
\mn{adom}(\Cmc) \times \{\mathsf{core},\mathsf{subtree}\}$. It will
accept this triple iff
\begin{enumerate}
\item there is a homomorphism $h$ from $q_t$ to the universal model of
$\Amc$ and $\Omc$ that maps answer variables of $q_t$ to $\abf$ and
$\lambda(y) = (h(y), \mathsf{core})$ if $h(y) \in \Csf_\mn{core}$ and $\lambda(y) = (c, \mathsf{subtree})$, if $h(y)$ maps into a fresh subtree in the universal model of $\Amc$ and $\Omc$ whose root is $c \in \Csf_\mn{core}$ and
\item $p \to q$.
\end{enumerate}

It is easy to see that Condition~1 holds for some $\lambda$ if and only if Condition~2$'$ from the definition of the UCQ-rewriting $q_r$ in the main part of the paper holds. Recall that Condition~2$'$ is the following requirement:
$$
\abf' \in \mn{cert}_Q(\Cmc' \cup \{ A(a) \mid \Amc',\Omc
 \models A(a),\; a \in \mn{dom}(\Cmc')\}).
$$
Hence Condition~1 guarantees that $(\Amc,\abf)$ is in the rewriting $q_r$. Together with
the second point and the conditions on the encoded tuples
$(\Amc,\abf,p)$ this then means that there is a $\lambda$ such that
$\Amf^\lambda$ accepts a tree encoding $(\Amc,\abf,p)$ iff $p$ is a
disjunct in $\Mbf^-(q_r)$ such that $p \to q$.

The automaton $\Amf^\lambda$ is constructed as the intersection of three automata $\Amf_\mn{proper}$, $\Amf^\lambda_1$ and $\Amf_2$, where the first checks if the input tree describes a proper tree, $\Amf^\lambda_1$ accepts trees that encode a triple that satisfies Condition~1 and $\Amf_2$ accepts trees that encode a triple that satisfies Condition~2. It is easy to define the automaton $\Amf_\mn{proper}$, we leave out the details.

\smallskip
\noindent
\textbf{Definition of $\Amf^\lambda_1$.} To define $\Amf^\lambda_1$, we first introduce some notation. Let $\mn{ROL}$ be the set of roles that appear in $\Omc$ or $\mn{sch}(\Mbf)$. Let $\mn{CN}$ the set of concept names that appear in $\Omc$ or $\mn{sch}(\Mbf)$ and let $\mn{tp} = 2^\mn{CN}$. Let $U$ be the set of all partial functions from the variables in $q_t$ to the set $\{\mn{core}, \mn{subtree}\}$. We define a relation $R \subseteq \mn{tp} \times U$ such that $(t, f) \in R$ iff there is a homomorphism $g$ from $q_t$ restricted to variables in the domain of $f$ to the universal model of $\Omc$ and $\{A(a) \mid A \in t\}$ such that $g(y) = a$ iff $f(y) = \mn{core}$. The relation $R$ can be computed in exponential time, since answering ontology mediated conjunctive queries with an $\ELHI$-TBox is in $\ExpTime$. \cite{DBLP:conf/jelia/EiterGOS08}

Given a pair $(t,f)$, compute the universal model of $\{A(a) \mid A \in t\}$ and $\Omc$ up to depth $|q_t|$ and check for the existence of a suitable homomorphism $g$ using brute force.

Let $\Amf_1 = (S, \delta, \Sigma_\varepsilon \uplus \Sigma_N, s_0, c)$ where
\begin{align*}
S = \{s_0\} &\uplus \{s^A_b \mid A \in \mn{CN} \text{ and } b \in \Csf_{\mn{core}}\}\\
& \uplus \{s^A_{r,b} \mid A \in \mn{CN} \text{ and } r \in \mn{ROL} \text{ and } b \in \Csf_{\mn{core}}\}\\
& \uplus \{s^A_r \mid A \in \mn{CN} \text{ and } r \in \mn{ROL}\}\\
& \uplus \{s^A \mid A \in \mn{CN}\}
\end{align*}
and $c(s) = 1$ for every $s \in S$, i.e. precisely the finite runs are
accepting. All states besides $s_0$ are used to check whether a certain fact $A(a)$ is entailed in the universal model of $\Amc$ and $\Omc$. Following Lemma \ref{lem:derivationtrees}, this can be done by checking the existence of an appropriate derivation tree. The state $s^A_b$ checks whether $\Amc, \Omc \models A(b)$.
The state $s^A_{r,b}$ checks whether we are in an $r$-child $d$ of $b$
such that $\Amc, \Omc \models A(d)$. The state $s^A_r$ checks whether
the current node $d$ is an $r$-child such that $\Amc, \Omc \models
A(d)$. The state $s^A$ checks whether for the current node $d$ we have
$\Amc, \Omc \models A(d)$.

We now describe the definition of the transition function $\delta$.
To define $\delta(s_0, l)$, where $l = (\Bmc,\abf,\mu) \in
\Sigma_\varepsilon$ we distinguish cases depending on whether the
$\Bmc$ and the $\abf$ encoded in $l$ is compatible with the homomorphism pattern
$\lambda$. That is we check whether the following two conditions are fulfilled:
\begin{itemize}
\item $\lambda$ maps the $i$-th answer variable $x_i$ from $q_t$ to $(a_i, \mn{core})$, where $a_i$ is the $i$-the element of $\abf$.
\item For every role atom $r(z_1,z_2)$ in $q_t$ such that $\lambda(z_i) = (b_i, \mn{core})$ we have that $r(b_1, b_2) \in \Bmc$.
\end{itemize}
If these conditions are not fulfilled then we set $\delta(s_0,l) =
\mn{false}$, meaning that the automaton $\Amf^\lambda_1$ immediately
rejects the input tree. If these conditions are fulfilled then
define $\delta(s_0, l)$ to be:
\begin{align}
&\bigwedge_{\substack{z \in \mn{var}(q_t) \\ \lambda(z)=(b, \mn{core})}} \bigwedge_{A(z) \in q_t} \langle 0 \rangle s^A_b \qquad \wedge \label{tran:start-1}\\
&\bigwedge_{\substack{Z \subseteq \mn{var}(q_t) \\ Z = \lambda^{-1}(\{b\} \times \{\mn{core}, \mn{subtree}\}) \\ Z \neq \emptyset}} \bigvee_{\substack{t \in \mn{tp} \\ (t,f) \in R \\ f = \pi_2 \circ \lambda |_Z}} \bigwedge_{A \in t} \langle 0 \rangle s^A_b \label{tran:start-2}
\end{align}
Here $\lambda |_Z$ denotes the restriction of $\lambda$ to the
variables in $Z$ and $\pi_2$ denotes the projection to the second
component. The conjunction in the first line makes sure that the
concept names needed for variables of $q_t$ that are mapped to
$\Csf_{\mn{core}}$ are derived. The second line assures that for all
variables of $q_t$ that are mapped to a subtree generated by
existential restrictions in the universal model of $\Amc$ and $\Omc$,
there is actually a type $t$ derived at the root $b \in
\Csf_{\mn{core}}$ that generates a suitable tree.

The following transitions are then used to check for derivations of
concept names.

For $l \in \Sigma_\varepsilon$, let:
\begin{align*}
\delta(s^A_b,l) &= \bigvee_{\Omc \models B_1 \sqcap \dots \sqcap B_n \sqsubseteq A} \langle 0 \rangle q^{B_1}_b \wedge \ldots \wedge \langle 0 \rangle s^{B_n}_b \;\;  \vee \\
& \bigvee_{\substack{\exists s.B \sqsubseteq A \in \Omc \\ \Omc
\models
r \sqsubseteq s}} \bigvee_{\substack{b' \in \Csf_{\mn{core}} \\ r(b,b') \in \Bmc}} \langle 0 \rangle s^B_{b'} \;\;  \vee \\
& \bigvee_{\substack{\exists s.B \sqsubseteq A \in \Omc \\ \Omc
\models
r \sqsubseteq s}} \bigvee_{1 \leq i \leq |\Omc| \cdot |q_t|} \langle i \rangle s^B_{r,b}
\end{align*}
For $l \in \Sigma_N$ and $\{r, b\} \subseteq l$ let:
\begin{align*}
\delta(s^A_{r,b}, l) &= \langle 0 \rangle s^A.
\end{align*}
For $l \in \Sigma_N$ and $r \in l$ let
\begin{align*}
\delta(s^A_r, l) &= \langle 0 \rangle s^A.
\end{align*}
For $l \in \Sigma_N$ with role $r \in l$ such that $l$ contains no
constant of $\Csf_{\mn{core}}$, let:
\begin{align*}
\delta(s^A,l) &= \bigvee_{\Omc \models B_1\sqcap \dots \sqcap B_n \sqsubseteq A} \langle 0 \rangle s^{B_1} \wedge \ldots \wedge \langle 0 \rangle s^{B_n} \;\; \vee \\
& \bigvee_{\substack{\exists s^-.B \sqsubseteq A \in \Omc \\ \Omc
\models r \sqsubseteq s}} \langle -1 \rangle s^B \;\; \vee \\
& \bigvee_{\substack{\exists s.B \sqsubseteq A \in \Omc \\ \Omc
\models u \sqsubseteq s}} \langle 1 \rangle s^B_u \vee \ldots \vee \langle |\Omc| \rangle s^B_u
\end{align*}
For $l \in \Sigma_N$ with $\{r,b\} \subseteq l$ for a role $r$ and $b
\in \Csf_{\mn{core}}$, let:
\begin{align*}
\delta(s^A,l) &= \bigvee_{\Omc \models B_1 \sqcap \dots \sqcap B_n \sqsubseteq A} \langle 0 \rangle s^{B_1} \wedge \ldots \wedge \langle 0 \rangle s^{B_n} \;\; \vee \\
& \bigvee_{\substack{\exists s^-.B \sqsubseteq A \in \Omc \\ \Omc
\models r \sqsubseteq s}} \langle -1 \rangle s^B_b \;\; \vee \\
& \bigvee_{\substack{\exists s.B \sqsubseteq A \in \Omc \\ \Omc
\models u \sqsubseteq s}} \langle 1 \rangle s^B_u \vee \ldots \vee \langle |\Omc| \rangle s^B_u
\end{align*}
All pairs $(s,l) \in S \times \Sigma_\varepsilon \uplus \Sigma_N$ that have not been mentioned yet will never occur in a run on a proper tree, so for those we just define $\delta(s,l) = \mn{false}$.

\smallskip
\noindent
\textbf{Correctness of $\Amf^\lambda_1$.} We now argue that $\Amf^\lambda_1$ accepts a tree if and only if it encodes a tuple $(\Amc, \abf, p)$ such that Condition 1 is fulfilled.

Let Condition~1 be fulfilled. We show that the automation accepts.
This entails that the homomorphism pattern $\lambda$ is compatible
with the $\Bmc$ and $\abf$ that encoded in the label $l$ at the root
node of the tree. Hence $\Amf^\lambda_1$ does not reject immediately.
Let $h$ be the homomorphism from Condition~1. Since $h$ is a
homomorphism for all variables $z$ of $q_t$ that are mapped to the
core of $\Amc$ and all atoms $A(z)$ from $q_t$, we have that $\Amc,
\Omc \models A(h(z))$, so the conjunction (\ref{tran:start-1}) will
succeed. For the second conjunction, consider a set of variables $Z
\subseteq \mn{var}(q_t)$ described by the first conjunction when
considering a core constant $b$. Let $t$ be the set of concept names
derived at $b$ in the universal model of $\Omc$ and $\Amc$. We argue
that that $(t,f) \in R$: By the definition of $\lambda$, the
homomorphism $h$ maps every variable from $Z$ either to $b$ or to the
subtree below $b$. Since $\Omc$ is assumed to be in normal form, the
subtree generated below $b$ only depends on $t$ and we can define $g$
to be the restriction of $h$ to $Z$. This function $g$ witnesses that
$(t,f) \in R$. For this set $t$, the last conjunction will of course
succeed, since $t$ was chosen to be the set of all concept names
derived at $b$.

For the other direction, let $(T, L)$ be a proper tree that is
accepted by $\Amf^\lambda_1$ and that encodes the tuple $(\Amc, \abf,
p)$. We need to construct the homomorphism $h$ such that Condition 1
is fulfilled. Since the automaton does not reject immediately and the
conjunction in (\ref{tran:start-2}) is fulfilled, there are $b_1,
\ldots, b_n \in \Csf_{\mn{core}}$ such that the sets $Z_i =
\lambda^{-1}(\{b_i\} \times \{\mn{core}, \mn{subtree}\}) \neq
\emptyset$ form a partition of $\mn{var}(q_t)$ and for every $i$ there
is a type $t_i$ derived at $b_i$ in the universal model of $\Omc$ and
$\Amc$ such that $(t_i,\pi_2 \circ \lambda |_{Z_i}) \in R$. The latter
means that there is a homomorphism $g_i$ from $q_t$ restricted to
$Z_i$ to the tree that is generated below $b_i$ in the universal model
of $\Omc$ and $\Amc$. The homomorphism $h$ is obtained by combining
the $g_i$ for all $i$. The second condition in the definition of
$\Sigma_\varepsilon$ guarantees that $h$ also preserves roles between
variables that lie in different $Z_i$.

\smallskip
\noindent
\textbf{Definition of $\Amf_2$.} To define $\Amf_2$, we precompute three relations $R$, $R'$ and $R''$. Let $R$ be a relation between ABoxes $\Bmc$ over $\Csf_{\mn{core}}$, disjuncts $d \in \Mbf^-(\Bmc)$ and functions $f$ from $\mn{adom}(\Bmc)$ to $\mn{var}(q)$. A triple $(\Bmc, d, f)$ is in $R$ if and only if there exists a homomorphism $h:d \to q$ such that $h(b) = f(b)$ for all $b \in \mn{adom}(\Bmc)$. Let $R'$ be a relation between unary mappings from $\Mbf$ and variables from $q$. A pair $(\varphi(x) \to A(x), y)$ is in $R'$ if and only if there is a homomorphism $h:\varphi(x) \to q$ such that $h(x) = y$. Let $R''$ be a relation between binary mappings from $\Mbf$ and pairs of variables from $q$. A triple $(\varphi(x,x') \to r(x,x'), y, y')$ is in $R''$ if there is a homomorphism $h : \varphi(x,x') \to q$ such that $h(x) = y$ and $h(x') = y'$. All three relations can be computed in $\ExpTime$, since they all only check for homomorphisms between structures of polynomial size.

Let $\Amf_2 = (S, \delta, \Sigma_\varepsilon \uplus \Sigma_N, s_0, c)$ where
\begin{align*}
S = \{s_0\} &\uplus \{s^b_y \mid b \in \Csf_{\mn{core}} \text{ and } y \in \mn{var}(q)\}\\
&\uplus \{s_y \mid y \in \mn{var}(q)\}
\end{align*}
and $c(s)=0$ for every $s \in S$, but the actual value of $c(s)$ does not matter since all runs of $\Amf_2$ on proper trees will be finite.

For $l \in \Sigma_\varepsilon$ and $\Bmc \in l$ and $d \in \Mbf^-(\Bmc)$ the disjunct defined by the mappings in $l$, we define:
\begin{align*}
\delta(s_0, l) &= \bigvee_{\substack{f : \mn{adom}(\Bmc) \to \mn{var}(q) \\ (\Bmc, d, f) \in R}} \bigwedge_{b \in \mn{adom}(\Bmc)} \bigwedge_{i \in \{1,\ldots,|\Csf_{\mn{core}}|\cdot|\Omc|\}} [i] s^b_{f(b)}
\end{align*}
For $l \in \Sigma_N$ we define:
\[
\delta(s^b_y, l) = \begin{cases}\mn{true} & \mbox{if } b \notin l \\ \langle 0 \rangle s_y & \mbox{if } b \in l\end{cases}
\]
For $l = (\Theta,M,\mu) \in \Sigma_N$ and $y \in \mn{var}(q)$ let
$Y^y_l$ be the set of all $y' \in \mn{var}(q)$ such that
$(\varphi(x,x') \rightarrow r(x,x'), y, y') \in R''$, where
$\varphi(x,x') \rightarrow r(x,x')$ is the mapping $M$ corresponding
to the unique role in $\Theta$, and such that $(\varphi(x) \to A(x),
y') \in R'$ for every concept name $A \in \Theta$, where $\varphi(x)
\to A(x)$ is the mapping $\mu(A)$. Then we define:
\[
\delta(s_y,l) \;=\; \bigvee_{y' \in Y^y_l} \; \bigwedge_{i \in \{1,\ldots,|\Omc|\}} [i] s_{y'}
\]

\smallskip
\noindent
\textbf{Correctness of $\Amf_2$.} We now argue that $\Amf_2$ accepts a tree $(T, L)$ if and only if it encodes a tuple $(\Amc, \abf, p)$ such that $p \to q$.

Assume there is homomorphism $h : p \to q$. We use $h$ to describe a run of $\Amf_2$ on $(T,L)$ that traverses the tree once from the root to the leaves. At the root, the automaton chooses as $f$ the restriction of $h$ to $\mn{adom}(\Bmc)$, which is possible because $(\Bmc, d, f) \in R$. If the run is at a position $(t,s_y)$, where $t \in T$ corresponding to a constant $a$ in $\Amc$, then we choose $h(a)$ as $y'$. Because $h$ is a homomorphism, one can check that $y' \in Y^y_{L(t)}$. Thus, $\Amc_2$ accepts $(T,L)$.

For the other direction, let $(T,L)$ be a tree encoding a tuple
$(\Amc, \abf, p)$ that is accepted by $\Amf_2$. Let $\rho$ be an
accepting run of $\Amf_2$ on $(T,L)$. We obtain a homomorphism $h:p
\to q$ by gluing together all of the following homomorphisms:
\begin{itemize}
\item The homomorphism from $p$ restricted to facts generated by facts
in $\Bmc$ to $q$  that exists by the choice of $f$ in the root node.
\item All homomorphisms $h'$ obtained as follows: Consider any
non-core fact $\alpha$ from $\Amc$. This fact appears in the label
$L(t)$ of some $t$ in $(T,L)$. Since $\rho$ is accepting, it will
visit $t$ in some state of the form $s_y$ and chooses $y' \in
Y^y_{L(t)}$. Because $\alpha$ is encoded in $L(t)$, it follows by the
definition of $Y^y_{L(t)}$ that there is a homomorphism from the body
of the mapping corresponding to $\alpha$ to $q$, which we choose as
$h'$.
\end{itemize}
The information that is passed along in the states of the automaton guarantees that all these homomorphisms can be glued together to obtain a single homomorphism $h:p \to q$.

\section{Lower Bounds for \EL, Missing Details}

We first give some justification of Theorem~\ref{lem:nexpbasic}.

\nexpbasic*
The hardness proof in \cite{BHLW-IJCAI16} actually produces as $q_2$ a
rooted CQ, but does not satisfy Condition~2. However, the only
exception to Condition~2 are CIs of the form $D \sqsubseteq C_{q}$
where $C_q$ is a specially crafted \ELI-concept that uses only symbols
from $\Sigma$ and is designed to `make
the query $q_2$ true', that is, whenever $d \in C_q^\Imc$ in an
interpretation \Imc, then $\Imc \models q_2(d)$. It can be verified that
the reduction in \cite{BHLW-IJCAI16} still works when replacing the rooted
CQ $q_2$ with the rooted UCQ $q_2 \vee \bigvee_{D \sqsubseteq C_{q}
  \in \Omc} q_D$, where $q_D$ is the concept $D$ viewed as a unary CQ
in the obvious way, and then deleting all CIs of the form $D
\sqsubseteq C_{q}$ from \Omc.  Arguably, this modification even yields
the more natural reduction, avoided in \cite{BHLW-IJCAI16} to ensure that
$q_2$ is a CQ rather than a UCQ.

Towards a proof of Lemma~\ref{lem:nexpCorrMain}, consider the OMQ
$Q=(\Omc',\Sbf,q_s)$ and the infinitary UCQ $q_r$ that consists of the
following CQs:
\begin{enumerate}

\item $B(x)$,

\item each CQ from $q_2(x)$,

%\item $M(x) \wedge q_{C_i}()$ for $1 \leq i \leq k$,

% \item for every $\Sigma$-ABox \Amc inconsistent with $\Omc$ and
%   $a \in \mn{Ind}(\Amc)$, $(\Amc,a)$ viewed as a CQ,

\item for every tree-shaped $\Sigma$-ABox \Amc %that is consistent
%                                with \Omc and has 
with root $a_0 \in \mn{cert}_{Q_1}(\Amc)$, $(\Amc,a)$ viewed
  as a CQ.

\end{enumerate}
The following is straightforward to verify. In particular, the
restriction to tree-shaped ABoxes in Point~3 is sanctioned
by Lemma~\ref{lem:unravel} and it is important that $q_2(x)$
uses only symbols from $\Sigma$ which cannot be derived
using the ontology.
\begin{lemma}
\label{lem:isrewr}
  $q_r$ is a rewriting of $Q$.  
\end{lemma}
The next lemma is the core ingredient to the proof of
Lemma~\ref{lem:nexpCorrMain}. In fact, by
Theorem~\ref{thm:maincharact} and since $\Mbf(q_s)=q_s$ and
$\Mbf^-(q_r)=q_r$, $q_s$ is UCQ-expressible in \Smc iff $q_r
\subseteq_\Sbf q_s$. The following lemma states that this is the
case iff $Q_1 \subseteq Q_2$.
\begin{lemma}
\label{lem:prepcorrlemnexp}
$Q_1 \subseteq Q_2$ iff for every CQ $p$ in $q_r$, there is a CQ $p'
\in q_s$ with $p' \rightarrow p$.
\end{lemma}

\noindent
\begin{proof}
  ``if''. Assume that $Q_1 \not\subseteq Q_2$. Then there is a
  $\Sigma$-ABox \Amc and an $a \in
  \mn{adom}(\Amc)$ such that $a \in \mn{cert}_{Q_1}(\Amc)$, but $a
  \notin \mn{cert}_{Q_2}(\Amc)$. By Lemma~\ref{lem:unravel} and as
  already observed in \cite{BHLW-IJCAI16}, we can assume that \Amc is
  tree-shaped with root $a$.  Let $q_{\Amc}$ be $(\Amc,a)$ viewed as
  a CQ. Clearly $a \in \mn{cert}_Q(\Amc)$ and thus $q_\Amc$ is a CQ
  in $q_r$. However, from none of the CQs in the UCQ $q_s$ there is a
  homomorphism to $q_{\Amc}$. This is true for $B(x)$ since
  $B$ does not occur in $q_{\Amc}$. It is also true for % each $M(x)
  % \wedge q_{C_i}()$ since otherwise \Amc would not be consistent with
  % \Omc. And finally it is true for
  $q_2(x)$ since $\Amc \not\models
  Q_2(a)$ and $q_2$ does not use symbols that occur on the right-hand
  side of CIs in \Omc.

  ``only if''. Assume that $Q_1 \subseteq Q_2$. We have to show that
  for every CQ $p$ in $q_r$, there is a CQ $p' \in q_s$ with $p'
  \rightarrow p$. This is clear for the CQs from Points~1 and~2 of the
  construction of $q_r$ since all of them appear as a CQ also in
  $q_s(x)$. For Point~3, let \Amc be a tree-shaped $\Sigma$-ABox with
  root $a_0 \in \mn{cert}_{Q_1}(\Amc)$. From $Q_1 \subseteq Q_2$, we
  obtain $a_0 \in \mn{cert}_{Q_2}(\Amc)$. Let $p$ be $(\Amc,a)$ viewed
  as a CQ.  By Points~1 and~2 from Theorem~\ref{lem:nexpbasic} , $a_0
  \in \mn{cert}_{Q_2}(\Amc)$ implies $a_0 \in \mn{ans}_{q_2}(\Amc)$ and
  thus $q_2 \rightarrow p$ and we are done.
\end{proof}
We now describe how the reduction can be improved to work for \EL,
that is, how the \ELI-ontology $\Omc'$ can be replaced
with an \EL-ontology. It can be verified that query $Q_1$ from
Theorem~\ref{lem:nexpbasic} is `one-way', that is, \Omc verifies the
existence of a (homomorphic image of a) certain tree-shaped sub-ABox
\emph{from the bottom up} and then $Q_1$ makes $q_1=A_0(x)$ true at
the root when the tree-shaped ABox was found. This one-way behaviour
can be made formal in terms of derivations of $A_0$ by \Omc, see
\cite{BHLW-IJCAI16}.

Assume w.l.o.g.\ that \Omc is in \emph{normal form}, that is, it only
contains CIs of the forms $\top \sqsubseteq A$, $A \sqsubseteq \bot$,
$A_1 \sqcap \cdots \sqcap A_n \sqsubseteq B$, $A \sqsubseteq \exists r
. B$, and $\exists r . A \sqsubseteq B$ where $A,B,A_1,\dots,A_n$ are
concept names.  A \emph{derivation} for a fact $A_0(a_0)$ in an
ABox \Amc with $A_0 \in \NC \cup \{ \bot \}$ is a finite
$\mn{adom}(\Amc) \times (\NC \cup \{\bot \})$-labeled tree $(T,V)$ that
satisfies the following conditions:
\begin{enumerate}

\item $V(\varepsilon)=(a_0,A_0)$;

% \item if $x \neq \varepsilon$, then $V(x)$ is not of the form
%   $(a,\bot)$;

% \item if $V(x)=(a,A)$ and $A(a) \in \Amc$ or $\top \sqsubseteq A \in
%   \Omc$, then $x$ is a leaf;

% \item if $V(x)=(a,A)$, then $x$ has one successor $y$ with $V(y)=(a,\Gamma)$
%   for some $\Gamma$ with $\Omc \models \Gamma \sqsubseteq A$;

\item if $V(x)=(a,A)$ and neither $A(a) \notin \Amc$ nor $\top
  \sqsubseteq A \in \Omc$, then one of the following holds:
  \begin{itemize}

  \item $x$ has successors $y_1,\dots,y_k$, $k \geq 1$ with
    $V(y_i)=(a,A_i)$ for $1 \leq i \leq k$ and $\Omc \models A_1
    \sqcap \cdots \sqcap A_k \sqsubseteq A$;
    % (then $x$ is of  \emph{local type});

  \item $x$ has a single successor $y$ with $V(y)=(b,B)$ and there is
an $\exists R . B \sqsubseteq A \in \Omc$ and a $R'(a,b) \in \Amc$
such that $\Omc \models R' \sqsubseteq R$;
    % (then $x$ is of \emph{existential type}).

  \item $x$ has a single successor $y$ with $V(y)=(b,B)$ and there
    is a $B \sqsubseteq \exists r . A \in \Omc$ such that $r(b,a) \in \Amc$ and
    $\mn{func}(r) \in \Omc$. 

  \end{itemize}
\end{enumerate}
Now if \Amc is tree-shaped, then the derivation is \emph{bottom-up} if
the following condition is satisfied: if $y$ is a successor of $x$
in~$T$, $V(x)=(a_x,A_x)$, and $V(y)=(a_y,A_y)$, then $a_x=a_y$ or
$a_y$ is a successor (but not a predecessor) or $a_x$ in \Amc, that
is, $a_y$ is further away from the root of \Amc than $a_x$ is. The OMQ
$Q_1$ is one-way in the sense that if \Amc is a tree-shaped
$\Sigma$-ABox with root $a_0 \in \mn{cert}_{Q_1}(\Amc)$, then all
derivations of $A_0(a_0)$ in \Amc are bottom-up. Note that a
corresponding statement for $Q_2$ makes little sense since by
Conditions~1 and~2 of Theorem~\ref{lem:nexpbasic}, answers to $Q_2$ on
an ABox \Amc are independent of \Omc.

We can exploit the one-way property of $Q_1$ as follows. In the
hardness proof in \cite{BHLW-IJCAI16}, all involved ontologies, signatures,
and queries use only a single role name $S$ that is interpreted as a
\emph{symmetric role}, and in fact represented via the composition
$r_0^-;r_0$ where $r_0$ is a fixed `standard' (non-symmetric) role
name. We can replace $S$ with a standard role name $r$ in \Omc and in
$\Sigma$, turning the $\ELI$-ontology \Omc into an $\EL$-ontology. The
mapping $r_0(x,y) \rightarrow r_0(x,y)$ from \Mbf in the original
reduction is then replaced with $r_0(x,y) \wedge r_0(x,z) \rightarrow
r(y,z)$; note that, in $q_s$, we keep the composition $r^-_0;r_0$ from
the original reduction.  We claim that, again, the following holds.
\begin{lemma}
  $Q_1 \subseteq Q_2$ iff $q_s$ is UCQ-expressible in \Smc.
\end{lemma}

\noindent
\begin{proof} (sketch) Let the UCQ $q_r$ be defined as before except
  that the CQs from $q$ are replaced with those from $\Mbf(q)$. It can
  be verified that $q_r$ is a rewriting of $Q=(\Omc',\Sbf,\Mbf(q_s))$.
  Moreover, it is easy to see that $\Mbf^-(\Mbf(q))=q$. This and the
  fact that $Q_1$ is one-way can be used to verify that $\Mbf^-(q_r)$
  is identical to the query $q_r$ from the original reduction (the
  one-way property is needed to see that the CQs from Point~3 of the
  definition of $q_r$ are identical in both cases, except that $S$ is
  replaced with $r$). We therefore get from
  Lemma~\ref{lem:prepcorrlemnexp} that
  \begin{description}

  \item[($*$)] $Q_1 \subseteq Q_2$ iff for every CQ $p$ in $\Mbf^-(q_r)$,
    there is a CQ $p' \in q_s$ with $p' \rightarrow p$.

  \end{description}
  By Theorem~\ref{thm:maincharact} $q_s$ is UCQ-expressible in \Smc
  iff $\Mbf^-(q_r) \subseteq_\Sbf q_s$. By ($*$), this is the case iff
  $Q_1 \subseteq Q_2$.
\end{proof}
Now for Point~2 of Theorem~\ref{thm:hardness}. We identify a
suitable containment problem proved 2\ExpTime in \cite{BHLW-IJCAI16}
and then proceed very similarly to the case of Point~1.
\begin{theorem}\cite{BHLW-IJCAI16}
\label{lem:2expbasic} 
Containment between OMQs $Q_1=(\Omc,\Sigma,q_1)$ and
$Q_2=(\Omc,\Sigma,q_2)$ with $\Omc$ an $\ELI$-ontology, $q_1$ of the
form $\exists x \, A_0(x)$, and $q_2$ a UCQ is 2\ExpTime-hard even
when
\begin{enumerate}

\item $q_2(x)$ uses only symbols from $\Sigma$ 

\item no symbol from $\Sigma$ occurs on the right-hand side of a CI in~\Omc;

\item all occurrences of $\bot$ in \Omc are of the form $C \sqsubseteq \bot$
  where $C$ is an \ELI-concept in signature $\Sigma$.

\end{enumerate}
\end{theorem}
Again, the actual hardness proof from \cite{BHLW-IJCAI16} needs to be
slightly modified to actually achieve what is stated in
Theorem~\ref{lem:2expbasic}. In particular, CIs of the form $D
\sqsubseteq C_q$ again have to be turned into additional disjuncts of
the UCQ $q_2$. This requires an additional modification of the
reduction since there are CIs of the form $D \sqsubseteq C_q$ where
$D$ uses symbols that are not from $\Sigma$ and occur on the
right-hand side of CIs in \Omc. In a nutshell and speaking in terms of
the notation from \cite{BHLW-IJCAI16}, the concept names $H$ and $W'$ need
to be added to $\Sigma$ and their presence at the intended places must
be `verified in the input' rather than `enforced by the
ontology'. After this is done, the only (minor) problem remaining is
that the concept name $G$ is used (exactly twice) in a (single) CQ $p$
in $q_2$, but it does occur on the right-hand side of two CIs which
are $G_1 \sqsubseteq G$ and $G_2 \sqsubseteq G$. Here, $G_1,G_2$ are
from $\Sigma$ and do not occur on the right-hand side of a CI. This
can be fixed by replacing $p$ with four CQs in the UCQ $q_2$,
replacing the two occurrences of $G$ with $G_1$ or $G_2$ in all
possible ways.

\smallskip

Point~2 of Theorem~\ref{thm:hardness} is again first proved for
$[\ELI,\text{GAV}]$ instead of for $[\EL,\text{GAV}]$. This is done by
reduction from the containment problem in Theorem~\ref{lem:2expbasic}.
Let $Q_1=(\Omc,\Sigma,A_0(x))$ and $Q_2=(\Omc,\Sigma,q)$.  We define
an OBDA-specification $\Smc = (\Omc',\Mbf,\Sbf)$ and a query $q_s()$
over \Sbf as follows. Let $B$ be a concept name that does not occur in
$Q_1$ and $Q_2$. Set
$$
\begin{array}{rcl}
  \Omc' &=& \Omc \cup \{ A_0 \sqsubseteq B \} \\[1mm]
  \Sbf &=& \Sigma \cup \{ B \}\\[1mm] 
  q_s() &=& \exists x \, B(x) \vee q()
\end{array}
$$
The set $\Mbf$ of mappings again contains $A(x) \rightarrow A(x)$ for
all concept names $A \in \Sbf$ and $r(x,y) \rightarrow r(x,y)$ for all
role names $r \in \Sbf$. The proof of the following is essentially
identical to the \coNExpTime case and omitted.
\begin{lemma}
\label{lem:2expCorrMain}
  $Q_1 \subseteq Q_2$ iff $q_s$ is UCQ-expressible in
  \Smc.
\end{lemma}
The approach to eliminating inverse roles is also exactly identical to
the \coNExpTime case. In fact, the OMQ $(\Omc,\Sigma,A_0(x))$ is again
one-way and all involved ontologies, signatures and queries again only
use the single symmetric role name $S$ represented as the composition
$r_0^-;r_0$. Consequently, the same arguments apply.

\section{Proof of Undecidability Result}

\thmalcfundec*

\noindent
\begin{proof}\
  We provide a reduction from the emptiness of AQs w.r.t.\
  \ALCF-ontologies, which is undecidable \cite{DBLP:journals/jair/BaaderBL16}. Thus
  let $(\Omc,\Sbf,A_0)$ be an OMQ with \Omc in \ALCF and $A_0(x)$ an
  AQ. Let $B_0$ be a fresh concept name and define an
  OBDA-specification $\Smc=(\Omc',\Mbf,\Sbf')$ where $\Omc' = \Omc
  \cup \{ A_0 \sqsubseteq B_0 \}$, $\Sbf'=\Sbf \cup \{ B_0 \}$, and
  \Mbf consists of the mappings $A(x) \rightarrow A(x)$ for all
  concept names $A$ in $\Sbf'$ and $r(x,y) \rightarrow r(x,y)$ for all
  role names $r$ in $\Sbf'$. We consider expressibility of the AQ
  $B_0(x)$. In fact, it is possible to verify the following:
  \begin{enumerate}

  \item if $A_0$ is empty w.r.t. \Omc, then $B_0(x)$ is a realization
    of $B_0(x)$ in \Smc;

  \item if $A_0$ is non-empty w.r.t.\ \Omc, then there is a $\Sbf$-database $D$ and a constant $a \in \mn{adom}(D)$ such that $a \in \mn{cert}_Q(D)$. Define the $\Sbf'$-database $D' = D \cup \{B_0(a)\}$.  
   Now, $B_0(x)$ is not
    determined in \Smc in the sense that
    $a \in \mn{ans}_{B_0}(D') \setminus \mn{ans}_{B_0}(D)$ but every
    model of $\Mbf(D)$ and $\Omc'$ is also a model of $\Mbf(D')$ and
    $\Omc'$, and vice versa. Consequently, $B_0(x)$ is not
    \Qmc-expressible in \Smc for any
    $\Qmc \in \{ \text{AQ}, \text{CQ}, \text{UCQ} \}$ or, in fact, any
    other query language.
  \end{enumerate}
\end{proof}

\fi

\end{document}